\documentstyle[amssymb,amsfonts,epsfig,12pt,cite,lscape]{article}
\newcommand{\lb}[1]{\label{#1}}

\newcommand{\Lra}{\Leftrightarrow}
\newcommand{\bc}{\begin{center}}
\newcommand{\ec}{\end{center}}
\newcommand{\be}{\begin{equation}}
\newcommand{\ee}{\end{equation}}
\newcommand{\bea}{\begin{eqnarray}}
\newcommand{\eea}{\end{eqnarray}}
\newcommand{\ba}[1]{\begin{array}{#1}}
\newcommand{\ea}{\end{array}}

\newcommand{\bt}[1]{\begin{table}[ht]\centering\begin{tabular}{#1}}
\newcommand{\et}[1]{\end{tabular}\caption{\small#1}\end{table}}

\newcommand{\sign}{\,{\mathrm{sign}}\,}

\newcommand{\diag}{{\mathrm{diag}}}

\begin{document}


\thispagestyle{empty}

\begin{flushright}

{\small
}

\end{flushright}

\begin{center}

{\large {\bf Rotating Magnetic Solutions for\\[2mm] 2+1D Einstein Maxwell Chern-Simons}}\\[10mm]

\vspace{1 truecm}

\vspace{1 truecm}
{\bf P. Castelo Ferreira}\\[6mm] {\small Centre for Rapid and Sustainable Product Development\\ Polytechnic Institute of Leiria}\\[6mm]
{\tt pedro.castelo.ferreira@gmail.com}
\ \\[25mm]

{\bf\sc Abstract\\[5mm]}
\begin{minipage}{15cm}
\vspace{2mm}
In this article are computed magnetic solutions of Einstein Maxwell Chern-Simons theory coupled to a dilaton-like scalar field. These solutions are computed by applying a space-time duality suggested by the author to known electric solutions of the same theory. As a redundancy check for the space-time duality it is explicitly shown that the magnetic configurations obtained are, as expected, solutions of the equations of motion. The magnetic configurations have metric determinant $\sqrt{-g}\sim r^p$ for the range of the parameter $p\in]-\infty,+\infty[/\{-1\}$ and are interpreted either as magnetic string-like configurations, configurations driven by an externally applied magnetic field or cosmological-like solutions with background magnetic fields.
\end{minipage}

\end{center}

\noindent

\vfill
\begin{flushleft}
\end{flushleft}

\newpage

\addtolength{\baselineskip}{0.20\baselineskip}

\thispagestyle{empty}
\tableofcontents

\newpage

\setcounter{page}{1}

\setcounter{equation}{0}
\section{Introduction\lb{sec.int}}

The first studies on classical gravitational solutions in $2+1$-dimensional
space-times date back to 1984 and addressed Cosmological Einstein theories~\cite{DJH_1,DJH_2}.
Later developments addressed neutral solutions for Einstein theory
(AdS BTZ black-hole)~\cite{BTZ_01}, Einstein Chern-Simons theory~\cite{Kogan1,Kogan2}
and the rotating BTZ black-hole~\cite{BTZ_02,Carlip1}. Following
these developments charged solutions were studied for
Einstein Maxwell Chern-Simons theory~\cite{CS_02,CS_03,CS_04,CS_05,CS_06},
Einstein Maxwell theory~\cite{BTZ_03,lemos_01},
Dilaton Einstein Maxwell theories~\cite{CM_01,CM_02,lemos_02,SLO_01,SLO_02,
SLO_03,SLO_04,SLO_05} and electric solutions of Einstein
Maxwell Chern-Simons theory with a scalar field~\cite{electric}, as well as
for Chern-Simons gravity~\cite{CS_grav_03,CS_grav_04,CS_grav_05,CS_grav_06,CS_grav_07,CS_grav_08,CS_grav_09}.

In this article are computed new magnetic solutions that further extend the known existing solutions for Einstein Maxwell Chern-Simons theory coupled to a Dilaton-like scalar field. As a motivation for the several fields and sectors of the full theory studied here it is relevant to note that, when considering $2+1$-dimensional gravitational solutions the inclusion of a scalar field is a natural extension of Einstein theory and it is justified by noting that a dimensional reduction from $3+1$-dimensions generates such a scalar field, whether it is a Dilaton field~\cite{CM_01,CM_02,dilpot}
or obtained by gauging a higher dimensional symmetry~\cite{cyl_01,cyl_02}. In addition when considering electromagnetic field solutions in $2+1$-dimensions
the Chern-Simons term~\cite{CS_grav_01,CS_grav_02} is also a natural extension of Maxwell theory, at quantum level only the Maxwell Chern-Simons theory is consistent such that the Chern-Simons term is a quantum correction of the Maxwell theory~\cite{qed3_00,qed3_01,qed3_02,dunne}.

As possible physical frameworks were such solutions may be relevant we note that $2+1$-dimensional
theories are often considered simpler laboratories for higher dimensional theories~\cite{Carlip1},
higher dimensional examples with similar frameworks to the one discussed here are: inflationary models
with exponential potentials~\cite{BB}; domain walls in 4+1-dimensions~\cite{branes_01}; and cosmological
solutions in 4+1-dimensions~\cite{5d_1,5d_2}. In addition often $3+1$-dimensional systems exhibiting 
cylindrical symmetry are considered as effective $2+1$-dimensional systems~\cite{cyl_01,cyl_02,lemos_03} as it
is the example of cylindrical gravitational waves~\cite{QG_01,QG_02,QG_03,QG_04,QG_05}.

To compute these new solutions it is applied the space-time suggested by the author in~\cite{duality} to the previously computed electric solutions for this theory~\cite{electric}. These dualities constitutes a generalization of a duality previously suggested in~\cite{Kogan1}. Shortly resuming the results obtained in~\cite{duality}, starting from a specific metric parameterization and Maxwell Chern-Simons Lagrangian
\be
\ba{rcl}
ds^2&=&-f^2dt^2+dr^2+h^2(d\varphi+Adt)^2\ ,\\[3mm]
{\mathcal{L}}^{\mathrm{MCS}}&=&F\wedge*F+m\,A\wedge F\ ,
\ea
\lb{ds}
\ee
with the standard electric and magnetic field definitions
\be
\ba{rcl}
E_*&=&F_{tr}=\partial_t A_r-\partial_r A_t\ ,\\[3mm]
B_*&=&F_{r\varphi}=\partial_r A_\varphi-\partial_\varphi A_r\ ,
\ea
\ee
where stared fields $E_*$ and $B_*$ stand for the electromagnetic fields in a specific coordinate
frame while non stared field $E$ and $B$ stand for the fields in the Cartan-frame (details are given in appendix~\ref{A.magnetic}),
there are three possible dualities that map electric into magnetic solutions. Specifically the interchange between time and angular variable corresponding to the two distinct duality maps
\be
\left\{\ba{rcl}
t&\to& i\varphi\\[3mm]
\varphi&\to& it
\ea\right.\ \Rightarrow\
\left\{\ba{rcl}
f&\to& ih\\[3mm]
h&\to& if
\ea\right.\ \ ,\ \
\left\{\ba{rcl}
E_*&\to& -iB_*\\[3mm]
B_*&\to& -iE_*\ ,
\ea\right.
\lb{duality1}
\ee
and
\be
\left\{\ba{rcl}
t&\to& \varphi\\[3mm]
\varphi&\to& t
\ea\right.\ \Rightarrow\
\left\{\ba{rcl}
f&\to& h\\[3mm]
h&\to& f
\ea\right.\ \ ,\ \
\left\{\ba{rcl}
E_*&\to& -B_*\\[3mm]
B_*&\to& -E_*\ .
\ea\right.
\lb{duality2}
\ee
A third duality map relates both these dualities~(\ref{duality1}) and~(\ref{duality2}) by a double Wick rotation
\be
\left\{\ba{rcl}
t&\to& it\\[3mm]
\varphi&\to& i\varphi
\ea\right.\ \Rightarrow\
\left\{\ba{rcl}
f&\to& if\\[3mm]
h&\to& ih
\ea\right.\ \ ,\ \
\left\{\ba{rcl}
E_*&\to& iE_*\\[3mm]
B_*&\to& iB_*\ .
\ea\right.
\lb{Wick_fields}
\ee

This work is organized as follows, in section~\ref{sec.BSolutions} the space-time dualities are applied to the electric gravitational solutions computed in~\cite{electric} which are in this way mapped into new magnetic gravitational solutions. Are also analyzed the singularities, curvature, horizons, mass, angular momentum and magnetic flux for these magnetic configurations.
In section~\ref{sec.conc} are summarized and discussed the results obtained, in particular
are interpreted either as magnetic string-like solutions, configurations driven by an external magnetic field
or cosmological-like solutions. In addition in appendix~\ref{A.magnetic} are re-derived directly from the
equations of motion in the Cartan-frame the solutions discussed in section~\ref{sec.BSolutions} and in appendix~\ref{sec.B} are listed, for particular cases not included in the main text, the expressions for the mass, angular momentum and magnetic flux.

\setcounter{equation}{0}
\section{Magnetic solutions for Minkowski space-time\lb{sec.BSolutions}}

In this section we derive explicit magnetic solutions for Einstein Maxwell Chern-Simons coupled to a scalar field
employing the space-time duality developed in the previous section applied to the electric solutions computed in~\cite{electric}.
Hence we are considering the same Action of~\cite{electric} that explicitly written in tensor notation is
\be
\ba{rcl}
S=&\displaystyle\frac{1}{2\pi}\int_M d^3x\left\{\sqrt{-\tilde{g}}\left[e^{a\phi}\left(\tilde{R}+2\lambda(\partial\phi)^2\right)-e^{b\phi}\Lambda\right.\right.\\[5mm]
&\displaystyle\left.\left.+\hat{\epsilon}\frac{e^{c\phi}}{2}\,\tilde{F}_{\mu\nu}\tilde{F}^{\mu\nu}\right]-\hat{\epsilon}\frac{m}{2}\,\epsilon^{\mu\nu\lambda}\tilde{A}_\mu \tilde{F}_{\nu\lambda}\right\}\ ,
\ea
\lb{S}
\ee
where $\hat{\epsilon}=\pm 1$ sets the relative sign between the gauge and gravitational sector
and the remaining terms follow the conventions of~\cite{electric} such that the metric has Minkowski ADM signature $\diag(-,+,+)$
and we are employing natural units $c_{\mathrm{light}}=\hbar=1$. We recall that $\hat{\epsilon}=+1$ stands for a ghost gauge sector such that the gauge fields contribution to the total energy is negative while $\hat{\epsilon}=-1$ stands for a standard gauge sector such that the
gauge fields contribution to the total energy is positive~\cite{gravitation,electric, duality}.

\subsection{Obtaining the solutions employing space-time duality\label{sec.solutions}}

Directly applying the duality map~(\ref{duality1}) to the electric solutions studied in~\cite{electric} accounts for mapping the metric parameterization and Maxwell Chern-Simons~(\ref{ds}) into
\be
\ba{rcl}
d\tilde{s}^2&=&-\tilde{f}^2(dt+\tilde{A}d\varphi)^2+dr^2+\tilde{h}^2d\varphi^2\ ,\\[4mm]
\tilde{{\mathcal{L}}}^{\mathrm{MCS}}&=&-\tilde{F}\wedge*\tilde{F}-m\tilde{A}\wedge \tilde{F}\ ,
\ea
\lb{gpar}
\ee
which is equivalent to the metric components map
\be
\left\{\ba{rcl}
\tilde{g}_{00}&=&-f^2+h^2A^2\\
\tilde{g}_{11}&=&1\\
\tilde{g}_{22}&=&h^2\\
\tilde{g}_{02}&=&h^2A
\ea\right.\ \ ,\ \
\left\{\ba{rcl}
\tilde{g}_{00}&=&-\tilde{f}^2\\
\tilde{g}_{11}&=&1\\
\tilde{g}_{22}&=&\tilde{h}^2-\tilde{f}^2\tilde{A}^2\\
\tilde{g}_{02}&=&-\tilde{f}^2\tilde{A}
\ea\right.\ ,
\lb{metric_map1}
\ee
and the field components map
\be
\left\{\ba{rcl}
f^2&=&\displaystyle \frac{\tilde{f}^2\,\tilde{h}^2}{\tilde{h}^2-\tilde{f}^2\,\tilde{A}^2}\\[3mm]
h^2&=&\displaystyle \tilde{h}^2-\tilde{f}^2\,\tilde{A}^2\\[3mm]
A&=&\displaystyle -\frac{\tilde{A}\,\tilde{f}^2}{\tilde{h}^2-\tilde{f}^2\,\tilde{A}^2}
\ea\right.\ \ ,\ \
\left\{\ba{rcl}
E_*&=&\displaystyle i\tilde{B}_*\left(\tilde{f}=f,\tilde{h}=h\right)\\[3mm]
B_*&=&\displaystyle i\tilde{E}_*\left(\tilde{f}=f,\tilde{h}=h\right)
\ea\right.\ .
\lb{map1}
\ee

Hence the magnetic solutions for the action~(\ref{S}) with $a=0$, $c=-b/2$ and $\lambda\neq b^2/8$~\cite{electric}
\be
\ba{rcl}
\phi&=&\displaystyle -\frac{2}{b}\ln(C_\phi\, r)\\[7mm]
\tilde{f}&=&\displaystyle C_f\,\sqrt{r}\\[7mm]
\tilde{h}&=&\displaystyle C_h\,r^{p-\frac{1}{2}}\\[7mm]
\tilde{A}&=&\displaystyle C_A\,r^{p-1}+\theta \\[7mm]
\tilde{B}_*&=&\displaystyle C_B\,r^{p-2}\\[7mm]
\tilde{A}_\varphi&=&\displaystyle \frac{C_B}{p-1}\,r^{p-1}
\ea
\lb{solsB}
\ee
where $C_h$, $C_f$, $b$ and $\theta$ are free parameters and the
constants $\lambda$, $C_\phi$, $C_A$ and $C_B$ have the following allowed values
\be
\ba{rcl}
\lambda&=&\displaystyle-\frac{b^2}{8}p\ ,\\[5mm]
C_\phi&=&\displaystyle|m|\sqrt{\frac{1-6x}{1-3p}}\ ,\\[5mm]
C_A&=&\displaystyle\frac{\sign(m)\,C_h}{C_f(1-p)}\sqrt{\frac{1-3p}{1-6x}}\ ,\\[5mm]
C_B&=&\displaystyle\frac{C_h}{\sqrt{2|m|}}\sqrt{\frac{\hat{\epsilon}(p-4x+6px)}{1-3p}}\left(\frac{1-3p}{1-6x}\right)^\frac{3}{4}\ ,
\ea
\lb{constantsB}
\ee
expressed in terms of a numerical parameter $p=p(x)$ and the ratio of the cosmological constant $\Lambda$ to the topological mass squared $m^2$
\be
x=\frac{\Lambda}{m^2}\ .
\ee
In the above expression for $C_B$ the factor of $(1-3p)$ was not simplified in order to maintain the factor inside of the square root explicitly
positive. For completeness the equations of motion in the Cartan-frame for this metric parameterization
are also solved in Appendix~\ref{A.magnetic}.

Imposing reality conditions for the solution constants there are four distinct allowed solutions depending on the parameter
$\hat{\epsilon}=\pm 1$, the range of values for the ratio $x$ and the respective bounds on the parameter $p$
\be
\ba{lll}
\mathrm{I.}&\hat{\epsilon}=+1\ ,\ &\displaystyle x\in\left]0,\frac{1}{2}\right]\ ,\\[4mm]
& &\displaystyle p=-\frac{3x-\sqrt{x(2-3x)}}{1-6x}\in\left]0,\frac{1}{2}\right]\\[6mm]
\mathrm{II.}&\hat{\epsilon}=-1\ ,\ &\displaystyle x\in\left]0,\frac{1}{6}\right[\,\cup\,\left[\frac{1}{2},\frac{2}{3}\right]\ ,\\[4mm]
& &\displaystyle p=-\frac{3x-\sqrt{x(2-3x)}}{1-6x}\in\left]0,\frac{1}{3}\right[\,\cup\,\left[\frac{1}{2},\frac{2}{3}\right]\\[6mm]
\mathrm{III.}&\hat{\epsilon}=+1\ ,\ &\displaystyle x\in\left]0,\frac{1}{6}\right[/\left\{\frac{1}{14}\right\}\ ,\\[4mm]
& &\displaystyle p=-\frac{3x+\sqrt{x(2-3x)}}{1-6x}\in\left]-\infty,0\right[/\{-1\}\\[6mm]
\mathrm{IV.}&\hat{\epsilon}=-1\ ,\ &\displaystyle x\in\left]\frac{1}{6},\frac{2}{3}\right]\ ,\\[4mm]
& &\displaystyle p=-\frac{3x+\sqrt{x(2-3x)}}{1-6x}\in\left[\frac{2}{3},+\infty\right[\ .
\ea
\lb{3sols}
\ee
The particular case $x=p=0$ corresponds to $\lambda=0$ and allows both for a limiting solution with $\tilde{B}_*=0$, $m\neq 0$, $\phi\neq 0$ (a non-trivial dilaton field) and the trivial solution with $m=\Lambda=\phi=\tilde{B}_*=0$ corresponding to empty flat Minkowski space-time. We note that the value of the cosmological constant is constrained by the mass $\Lambda<m^2$~(\ref{A.final_exp}) such that either of the limits
$\Lambda\to 0$ or $m\to 0$ are equivalent to the limit $x\to 0$. In the following we consider that the particular case $x\to 0$ is retrieved by taking the limit $m\to 0$ such that this limiting solution corresponds to the trivial solution, empty flat Minkowski space-time. In solution I, the particular case $x=1/6$ is well defined corresponding to the same solutions~(\ref{solsB}) with $p=1/3$ however for solution II this value of the parameter does not allow for a real solution. Both in solution I and II the parameter value $x=p=1/2$ is a well defined solution with null
magnetic field, $C_B=0$. In solution II and IV the parameter value $x=p=2/3$ is also a well defined solution. In solution III the particular case $p=-1$ corresponding to $x=1/14$ ($\lambda=b^2/8$) does not allow for solutions of the equations of motion, hence this value of the parameter is excluded. In solutions III and IV the value of the parameter $x=1/6$ corresponds to $-\infty$ and $+\infty$, respectively. In addition, for solution IV, the particular case $x=1/2$ corresponding to $p=1$ has the solutions for $h$, $f$, $\tilde{B}_*$ and $\phi$ given in~(\ref{solsB}) and~(\ref{constantsB}), however it has the particular solution for $A$
\be
p=1\,\Rightarrow\,\tilde{A}=C_A\,\log(r)+\theta\ \ ,\ \ C_A=\frac{C_h\sign(m)}{C_f}.
\lb{A_p1}
\ee

All the solutions presented correspond to positive cosmological constant and the solutions~I,~II with $x\in]0,1/6[$ and~III allow for the limiting solution corresponding to empty flat Minkowski space-time, $x\to 0$, while solution~II with $x\in[1/2,2/3]$ and solution~IV do not allow to obtain this limiting solution. For these solutions, the line element~(\ref{gpar}), re-written for the standard ADM parameterization~(\ref{ds}),
is $d\tilde{s}^2=-f^2dt^2+dr^2+h^2(d\varphi+Adt)^2$ with
\be
\ba{rcl}
f^2&=&\displaystyle \frac{C_f^2\,r^{2p-1}}{1-r^{2p-2}\,\left(\tilde{C}_A\,r^{p-1}+\tilde{\theta}\right)^2}\\[3mm]
h^2&=&\displaystyle C_h^2\,r\,\left(1-r^{2p-2}\,\left(\tilde{C}_A\,r^{p-1}+\tilde{\theta}\right)^2\right)\\[3mm]
A&=&\displaystyle -\frac{C_f\,r^{2p-2}\,\left(\tilde{C}_A\,r^{p-1}+\tilde{\theta}\right)}{C_h\left(1-r^{2p-2}\,\left(\tilde{C}_A\,r^{p-1}+\tilde{\theta}\right)^2\right)}
\ea
\lb{fhAtilde}
\ee
This metric has determinant $\sqrt{-g}=|C_f\,C_h|\,r^{p}$ and this parameterization is obtained directly from the map~(\ref{map1}) corresponding to the duality~(\ref{duality1}). The non-null metric components can be computed directly from this parameterization as expressed in equation~(\ref{metric_map1})
\be
\ba{rcl}
\tilde{g}_{00}&=&-C_f^2\,r^{2p-1}\ ,\\[5mm]
\tilde{g}_{11}&=&1\ ,\\[5mm]
\tilde{g}_{22}&=&C_h^2\,r\,\left(1-r^{2p-2}\,(\tilde{C}_A\,r^{p-1}+\tilde{\theta})^2\right)\ ,\\[5mm]
\tilde{g}_{02}&=&-C_f\,C_h\,r^{2p-1}\,(\tilde{C}_A\,r^{p-1}+\tilde{\theta})\ .
\ea
\lb{g_tilde}
\ee
In the above expressions we have replace the constants $C_A$~(\ref{constantsB}) and $\theta$ by the respective expressions multiplied by the ratio $C_f/C_h$
\be
\tilde{\theta}=\frac{C_f}{C_h}\,\theta\ \ ,\ \ \tilde{C}_A=\frac{C_f}{C_h}\,C_A=\displaystyle\frac{\sign(m)}{(1-p)}\sqrt{\frac{1-3p}{1-6x}}\ .
\lb{CA_tilde}
\ee

We note that for $p<1$ the metric has ADM signature $\diag(-,+,+)$ corresponding to the chosen convention
while for $p>1$ the metric has ADM signature $\diag(+,+,-)$ such that further considering a radial coordinate transformation $r\to 1/r$
it is obtained the metric ADM signature $\diag(+,-,-)$, hence corresponding to the opposite convention with respect to the originally chosen convention. At $p=1$ the metric ADM signature depends on the sign of the factor $(1-(\tilde{C}_A+\tilde{\theta})^2)$, when this factor is positive it has ADM signature $\diag(-,+,+)$ and when this factor is negative (considering the coordinate transformation $r\to 1/r$) it has ADM signature $\diag(+,-,-)$. We recall that the coordinate transformation $r\to 1/r$ implies exchanging the origin with spatial infinity $r:0\leftrightarrow+\infty$.
In addition, when horizons are present this swapping of signature is equivalent to swapping the exterior region with the interior region
of the horizons.

For the solutions discussed here, the swapping of the metric ADM signature with respect to the chosen convention corresponds to solution IV with the parameter $x$ in the range $x\in]1/6,1/2]$. Generally, for a given particular solution changing the metric ADM signature, the duality corresponding to a double Wick rotation of the coordinates $t$ and $\varphi$~(\ref{Wick_fields}) could generate new solutions which would maintain the metric ADM signature. However for the solutions just computed, when considering the reality conditions on the fields discussed in appendix~\ref{sec.solutions}, the duality~(\ref{Wick_fields}) simply swaps the sign of $\hat{\epsilon}$ and the parameter $p$, hence no new solutions are obtained, instead solutions I and II are swapped with each other and solutions III and IV are swapped with each other.

\subsection{Singularities and curvature analysis\label{sec.curvature}}

In this section we analyze the space-time singularities, the existence of horizons and its location.

To analyze space-time singularities in $2+1$-dimensions it is enough to analyze the contraction of the Ricci scalar $R_{\mu\nu}$
with itself~\cite{electric}. For the solutions computed in the previous section this contraction is
\be
\ba{rcl}
R_{\mu\nu}R^{\mu\nu}&=&\displaystyle\frac{1}{4r^{12}}\left[\left(3-16\,p+34\,p^2-28\,p^3+8\,p^4\right)\,r^8\right.\\[5mm]
& &\displaystyle \left.+ 3\tilde{C}_A^4\,\left(p-1\right)^4\,r^{8p} - 2\tilde{C}_A^2(p-1)^2(4p-5)\times\right.\\[5mm]
&&\displaystyle\hfill\ \left.\times(4p-3)\,r^{4+4p}\right]\ .
\ea
\lb{RR}
\ee
For the particular case $p=1$ we obtain $R_{\mu\nu}R^{\mu\nu}=3/(2r^4)$, hence for $p\geq 1$ the dominant divergent term near $r=0$ is proportional to $\sim 1/r^4$, while for $p<1$ it is proportional to $~1/r^{12-8p}$ such that we conclude that there is a space-time singularity at $r=0$ for all values of $p$. In addition, for $p>3/2$ corresponding to $x<9/26$ in solution IV, spatial infinity is also a space-time singularity as the dominant divergent term is proportional to $\sim r^{8p-12}$.

As for the curvature it is
\be
R=\frac{-(1-6p+4p^2)\,r^4+\tilde{C}_A^2\,(p-1)^2\,r^{4p}}{2r^6}\ .
\lb{R}
\ee
For the particular case $p=1$ the curvature is $R=1/r^2$.
Consistently with the singularity analysis discussed above, for $p<3/2$ the curvature vanishes at spatial infinity, hence space-time is
asymptotically flat, for $p=3/2$ it converges to the positive constant $\tilde{C}_A^2/8$ and for $p>3/2$ it diverges.

Depending on the values of $x$ the curvature is either always positive or exist regions where it is negative.
For solution I and II, it is always positive for $x\ge(8-3\sqrt{5})/38$, while for $x<(8-3\sqrt{5})/38$ it is negative
for $r>r_{0.\mathrm{I}}$ having a negative minimum value at $r=r_{\mathrm{min.I}}>r_{0.\mathrm{I}}$ and converging to 0 at spatial-infinity. Near the origin, for $r<r_{0.\mathrm{I}}$, it is positive. Here $r_{0.\mathrm{I}}$ and $r_{\mathrm{min.I}}$ are
\be
\ba{rcl}
r_{0.\mathrm{I}}&=&\displaystyle \left(1-7x-3\sqrt{x(2-3x)}\right)^{\frac{1}{4\left(p-1\right)}}\\
&&\hspace{5cm}\in\ ]1,+\infty[\ ,\\[6mm]
r_{\mathrm{min.I}}&=&\displaystyle \left(\frac{3-45x+102x^2-(7-22x)\sqrt{x(2-3x)}}{9-26x}\right)^{\frac{1}{4\left(p-1\right)}}\\
&&\hspace{5cm}\in\ ]3^{1/4},+\infty[\ .
\ea
\label{r0_rmin_I}
\ee
For solution III the curvature is negative for $r>r_{0.\mathrm{III}}$ having a negative minimum value at $r=r_{\mathrm{min.III}}>r_{0.\mathrm{III}}$, it converges to $0$ at spatial-infinity and near the origin, for $r<r_{0.\mathrm{III}}$, it is positive. As for solution IV, for $x\in]1/6,9/26[$ ($x=9/26$ corresponds to $p=3/2$), the curvature is negative for $r<r_{0.\mathrm{III}}$ and it is positive for $r>r_{0.\mathrm{III}}$ diverging at spatial infinity, for $x=9/26$ the curvature is negative for $r<2/\sqrt{13}$ and it is positive for $r>2/\sqrt{13}$ converging to $13/8$ at spatial infinity, for $x\in]9/26,(8+3\sqrt{5})/38[$ it is negative near the origin for $r<r_{0.\mathrm{III}}$ and it is positive for $r>r_{0.\mathrm{III}}$ converging to 0 at spatial infinity and it has a positive maximum value at $r=r_{\mathrm{min.III}}>r_{0.\mathrm{III}}$, while for $x\in[(8+3\sqrt{5})/38,2/3[$ the curvature is always positive converging to 0 at spatial infinity. Here $r_{0.\mathrm{III}}$ and $r_{\mathrm{min.III}}$ are
\be
\ba{rcl}
r_{0.\mathrm{III}}&=&\displaystyle \left(1-7x+3\sqrt{x(2-3x)}\right)^{\frac{1}{4\left(p-1\right)}}\\[3mm]&&\hspace{2cm}\in\ ]0.93,1[\ {\mathrm{for}}\ x\in\ ]0,1/6[ \\
&&\hspace{2cm}\in\ ]0,1.01[\ {\mathrm{for}}\ x\in\ ]1/6,2/3[\\[6mm]
r_{\mathrm{min.III}}&=&\displaystyle \left(\frac{3-45x+102x^2+(7-22x)\sqrt{x(2-3x)}}{9-26x}\right)^{\frac{1}{4\left(p-1\right)}}\\[3mm]&&\hspace{2cm}\in\ ]1,3^{1/4}[\ {\mathrm{for}}\ x\in\ ]0,1/6[ \\
&&\hspace{2cm}\in\ ]0,1[\ {\mathrm{for}}\ x\in\ ]9/26,(8+3\sqrt{5})/38[
\ea
\label{r0_rmin_III}
\ee

Hence, resuming the previous analysis, the curvature values for the several allowed solutions are, for the several solutions discussed, 
\be
\hspace{-2cm}
\ba{lll}
\mathrm{I.}&\hat{\epsilon}=+1 ,\\[4mm] &\displaystyle x\in\left]0,\frac{8-3\sqrt{5}}{38}\right[\ ,\ &R\in\left]R(r_{\mathrm{min.I}})<0,+\infty\right[\\[6mm]
                      &\displaystyle x\in\left[\frac{8-3\sqrt{5}}{38},\frac{1}{2}\right]\ ,\ &\displaystyle R\in\left]0,+\infty\right[\\[6mm]
\mathrm{II.}&\hat{\epsilon}=-1\ ,\\[4mm] &\displaystyle x\in\left]0,\frac{8-3\sqrt{5}}{38}\right[\ ,\ &R\in\left]R(r_{\mathrm{min.I}})<0,+\infty\right[\\[6mm]
                      &\displaystyle x\in\left[\frac{8-3\sqrt{5}}{38},\frac{1}{6}\right[\cup\left[\frac{1}{2},\frac{2}{3}\right] \ ,\ &\displaystyle R\in\left]0,+\infty\right[\\[6mm]
\mathrm{III.}&\hat{\epsilon}=+1\ ,\\[4mm] &\displaystyle x\in\left]0,\frac{1}{6}\right[/\left\{\frac{1}{14}\right\}\ ,\ &\displaystyle R\in\left]R(r_{\mathrm{min.III}})<0,+\infty\right[\\[6mm]
\mathrm{IV.}&\hat{\epsilon}=-1\ ,\\[4mm] &\displaystyle x\in\left]\frac{1}{6},\frac{9}{26}\right[\ ,\ &\displaystyle R\in\left]-\infty,+\infty\right[\\[6mm]
                      &\displaystyle x=\frac{9}{26}\ ,\ &\displaystyle R\in\left]-\infty,\frac{13}{8}\right[\\[6mm]
                      &\displaystyle x\in\left]\frac{9}{26},\frac{8+3\sqrt{5}}{38}\right[\ ,\ &R\in\left]-\infty,R(r_{\mathrm{min.III}})>0\right[\\[6mm]
                      &\displaystyle x\in\left[\frac{8+3\sqrt{5}}{38},\frac{2}{3}\right]\ ,\ &\displaystyle R\in\left]0,+\infty\right[
\ea
\lb{Rsols}
\ee

As for the nature of the space-time singularity we note that, independently of the value of the parameter
$p$, the maximum value of the coordinate $\varphi$ diverges at the singularity $r=0$ and it is finite up to spatial infinity being real outside
the horizon~(\ref{r_H}) discussed in the next section. At spatial infinity it diverges for $p\leq -1/4$ and $p>3/2$ and it is asymptotically null for $p\in[1,3/2[$ (being finite at $p=3/2$). As for the range of the parameter $p\in]-1/4,1[$ there is a specific
frame for which the maximum value of the coordinate $\varphi$ matches the usual relations corresponding to flat Minkowski space-time.
Specifically, defining the 2-dimensional intrinsic metric $\tilde{h}_{ij}=\diag(1,h^2)$ corresponding to metric $\tilde{g}_{\mu\nu}$~(\ref{g_tilde}) and considering a rescaling of the radial coordinate
\be
r=\tilde{r}^\xi\ \Rightarrow\ dr=\xi\,\tilde{r}^{\xi-1}\,d\tilde{r}\ ,
\ee
we obtain that the maximum value for the coordinate $\varphi$ is
\be
\ba{rcl}
\varphi_{\mathrm{max}}&=&\displaystyle\frac{2\pi}{\sqrt{-|g_{\mu\nu}|}}\sqrt{\frac{h_{\varphi\varphi}}{h_{rr}}}=\frac{2\pi}{f\,\sqrt{\,h_{rr}}}\\[5mm]
&=&\displaystyle\frac{2\pi\tilde{r}^{1-\xi(2p+1/2)}}{|\xi\,C_f|}\sqrt{1-\tilde{r}^{2\xi(p-1)}\left(\tilde{C}_A\tilde{r}^{\xi(p-1)}+\tilde{\theta}\right)^2}\ .
\ea
\lb{varphi_max}
\ee
such that the following asymptotic expressions at spatial infinity are obtained
\be
\ba{rcl}
\displaystyle\sqrt{-|\tilde{g}_{\mu\nu}|}&=&\displaystyle|\xi\,C_fC_h|\,\tilde{r}^{-1+\xi(1+2p)}\ ,\\[5mm]
\displaystyle\lim_{r\to\infty}\sqrt{|h_{ij}|}&=&\displaystyle|\xi\,C_h|\,\tilde{r}^{-1+\frac{3}{2}\xi}\ ,\\[5mm]
\displaystyle\lim_{r\to\infty}\varphi_{\mathrm{max}}&=&\displaystyle\frac{2\pi}{|\xi\,C_f|}\,\tilde{r}^{1-\frac{\xi(1+4p)}{2}}\ .
\ea
\lb{phimax_limits}
\ee
Setting $\xi=2/(1+4p)$, $\varphi_{\mathrm{max}}$ is asymptotically constant exactly matching $2\pi$ for $C_f=(1+4p)/2$. In addition we note that at spatial infinity both the space-time measure $\sqrt{-|g_{\mu\nu}|}$ and the space measure $\sqrt{|h_{ij}|}$ are, in this frame proportional to a positive exponent of $\tilde{r}$. Let us further note that the constant $C_f$ is interpreted as the velocity of light in vacuum and its value can be redefined by a re-scaling of the time coordinate $t$, hence there is some loss of generality when fixing the constant $C_f=(1+4p)/2$ (to ensure that $\lim_{\tilde{r}\to+\infty}\varphi=2\pi$) as we are fixing the speed of light in a particular frame, hence we are generally leaving $C_f$ as a free constant. Resuming this discussion we conclude that the coordinate $\varphi$ can exactly match the angular coordinate for Minkowski empty flat space-time at spatial infinity for a particular frame only when
\be
p\in\displaystyle\left]-\frac{1}{4},\frac{1}{2}\right[/\{0\}\ ,\\[5mm]
\lb{p_compact}
\ee 
When considering this constraint the range of the parameter $x$ for solution I is not affected, for solution II is reduced to $x\in]0,1/6[$, for solution III is reduced to $x\in]0,1/62[$ and solution IV is excluded. We further note that for this range only the space-time singularity at the origin exists~(\ref{singularities}) as $p<3/2$ such that no singularity at spatial infinity is present.

For all values of $p$, $\varphi_{\mathrm{max}}$ diverges at the singularity $r=0$ and, for $p>3/2$, $\varphi_{\mathrm{max}}$ is null at the singularity $r\to+\infty$, hence we interpreted these singularities as a decompactification singularity and a conical singularity, respectively~\cite{electric}
\be
\hspace{-2cm}\ \ba{rcl}
\displaystyle\forall_p\ ,\ \lim_{r\to 0}\varphi_{\mathrm{max}}=+\infty&\Rightarrow&r=0\ \mathrm{is\ a\ decompactification\ singularity}\ .\\[5mm]
\displaystyle p>\frac{3}{2}\ ,\ \lim_{r\to +\infty}\varphi_{\mathrm{max}}=0&\Rightarrow&r\to +\infty\ \mathrm{is\ a\ conical\ singularity}\ .
\ea
\lb{singularities}
\ee

Next we analyze the horizons for an external observer.

\subsection{Horizons and photon topological mass\label{sec.horizons}}

To analyze the existence of horizons the usual approach is to compute the geodesic motion of photons. From the point of view of an external
observer the horizon corresponds to the spatial hyper-surface for which the photon freezes such that its geodesic equation is $\dot{r}=0$.
In~\cite{electric} were computed the differential equations describing geodesic motion. For a particle with null angular momentum $L=0$ we obtain
\be
\begin{array}{rcl}
\dot{r}_{\kappa}&=&\displaystyle\pm\frac{\sqrt{-g\left(-g\frac{\kappa}{E^2}+g_{22}\right)}}{g_{22}}\\[3mm]
&=&\displaystyle \pm\frac{|C_f|r^{p-\frac{1}{2}}\sqrt{1+\frac{\kappa}{E^2}\,C_f^2\,r^{2p-1}-\,r^{2p-2}\,(\tilde{C}_A\,r^{p-1}+\tilde{\theta})^2}}{1-r^{2p-2}\,(\tilde{C}_A\,r^{p-1}+\tilde{\theta})^2}\ ,\\[5mm]
\dot{\varphi}_{\kappa}&=&\displaystyle-\frac{g_{02}}{g_{22}}=-\frac{C_f}{C_h}\times\frac{r^{2p-2}(\tilde{C}_A\,r^{p-1}+\tilde{\theta})}{1-\,r^{2p-2}(\tilde{C}_A\,r^{p-1}+\tilde{\theta})^2}\ ,
\end{array}
\lb{geod_eq}
\ee
where $g=|g_{\mu\nu}|$ is the determinant of the metric, $E$ the energy of the particle and $\kappa=-1$ for standard massive particles (corresponding to time-like trajectories), $\kappa=0$ for photons or any other massless particles (corresponding to light-like trajectories) and $\kappa=+1$ for tachyons or other particles with imaginary energy eigenvalues (corresponding to space-like trajectories).

Generally the above equations are not solvable analytically. In the following we will analyze the zeros and
divergences of the first equation for particles traveling towards the singularities which is enough to conclude whether a horizon exist
or not. We further note that due to the
Chern-Simons term the photon acquires a topological mass $m$ such that its energy squared is $E^2=m^2$~\cite{CS_grav_02}.
Specifically from the equation of motion for $A_\mu$ we obtain~\cite{electric} $\partial_\alpha(\sqrt{-g}\,e^{c\phi}F^{\alpha\mu})+m\epsilon^{\mu\alpha\beta}F_{\alpha\beta}/2=0$ such that computing the divergence
of this equation, replacing itself in the resulting differential equation and using the definition of the dual field strength
$*F^{\mu}=-\sqrt{-g}\,e^{c\phi}\epsilon^{\mu\alpha\beta}F_{\alpha\beta}/(2\sqrt{-g}\,e^{c\phi})$ we obtain the photon propagation equation in dual form~\cite{CS_grav_02}
\be
\left(\Box-m^2\right)*F^{\mu}=0\ \ ,\ \ \Box(\cdot)=\frac{1}{\sqrt{-g}\,e^{c\phi}}\partial_\alpha\left(\sqrt{-g}\,e^{c\phi}\partial^{\alpha}\left(\cdot\right)\right)\ ,
\lb{photon_mass}
\ee
where $\Box$ stands for the $2+1$-dimensional Laplace operator for action~(\ref{S}) and the relative signs in this equation do depend on the metric signature convention.
In particular we note that in flat space-time, for the convention adopted here, $\eta_{\mu\nu}\sim\diag(-,+,+)$ we consistently obtain $(-\partial_0\partial^0+\partial_i\partial^i-m^2)*F^\mu=0$ while for the opposite sign convention~\cite{CS_grav_02} $\eta_{\mu\nu}\sim\diag(+,-,-)$
we obtain $(\Box+m^2)*F^\mu=(\partial_0\partial^0-\partial_i\partial^i+m^2)*F^\mu=0$ such that both equations are the same up to an overall
minus sign, corresponding to a photon with a standard (topological) mass $m$. Hence as
extensively analyzed in the literature we conclude that no massless photons exist for Maxwell Chern-Simons theories~\cite{CS_grav_01,CS_grav_02}.

It is straight forward to check that for all values of $p$ and $\kappa$, as we approach the singularity at $r=0$, the velocity of any given
particle vanishes
\be
\lim_{r\to 0}\dot{r}_{\kappa}=0\ ,
\lb{hor_0}
\ee
while in this limit $\dot{\varphi}_{\kappa}$ is finite for $p=1$ and null for all other values of $p$.
This implies that the singularity is itself an horizon, hence it is not a naked singularity. However this result is not
conclusive as for higher values of $r>0$ there exists a divergence of $\dot{r}_{\kappa}$, specifically when the denominator of
the first equation of~(\ref{geod_eq}) is null the particle velocity diverges. This divergence is located at the value of the radial coordinate $r=r_{\mathrm{div}}$ obeying the equation
\be
1=\left(r_{\mathrm{div}}^{p-1}\,(\tilde{C}_A\,r^{p-1}_{\mathrm{div}}+\tilde{\theta})\right)^2\ .
\lb{r_div}
\ee
This equation has one real positive solution $r_{\mathrm{div}}$ for all values of $p$ and $\tilde{\theta}$. We recall that $\tilde{C}_A$ is not a free constant being expressed in equation~(\ref{constantsB}) and~(\ref{CA_tilde}) as a function of $x$ and $p=p(x)$. Specifically, one
of the 4 solution $r_{\mathrm{div},\pm,\pm}=((-\tilde{\theta}\pm\sqrt{\tilde{\theta}^2\pm 4\tilde{C}_A})/(2\tilde{C}_A))^\frac{1}{p-1}$, is real and positive for all the allowed range of the parameters.

In addition to ensure that for $r>r_{\mathrm{div}}$, the space-time has Minkowski signature $\diag(-,+,+)$ and that $\dot{r}_{\kappa}$ describes the  geodesic motion of a particle it is required that this quantity ($\dot{r}_{\kappa}$) be real valued and consistently
have either positive sign for particles traveling away from the singularity either negative sign for particles traveling towards the singularity. These properties are obeyed as long as the factor $1-\left(r^{p-1}\,(\tilde{C}_A\,r^{p-1}+\tilde{\theta})\right)^2$ is real and positive for $r>r_{\mathrm{div}}$. This statement is simply equivalent to the bound $p<1$ such that the factor $\left(r^{p-1}\,(\tilde{C}_A\,r^{p-1}+\tilde{\theta})\right)^2$ decreases with growing radial coordinate. Hence we obtain the bounds
\be
\left\{\ba{rcl}r&>&r_{\mathrm{div}}\\1&>&\displaystyle\left(r^{p-1}\,(\tilde{C}_A\,r^{p-1}+\tilde{\theta})\right)^2\ea\right.\Leftrightarrow p<1\ .
\lb{bound_p}
\ee
This bound, $p<1$, is consistent with the analysis in the previous section.

For massless particles the velocity divergence in $\dot{r}_{\kappa=0}$ just analyzed is outside any horizon. This is straight forwardly shown by noting that for $\kappa=0$ the numerator of $\dot{r}_{\kappa=0}$ is the square root of its denominator~(\ref{geod_eq}) such that the only horizon is at $r=0$ as already concluded~(\ref{hor_0}). Classically there is no interpretation for a particle velocity divergence, however we note that upon path integral quantization this phenomena can be consistently described as a tunneling effect, hence an instanton configuration~\cite{CS_grav_05}. We are not proceeding with this analysis here, instead let us note that from the photon equations of motion~(\ref{photon_mass}) the photon acquires a topological mass $m$ such that no massless photons exist in the theory discussed here. Therefore, assuming that no massless particles exist in the theory let us analyze the photon geodesic motion with energy squared given by $E^2=m^2$ and light-like trajectories ($\kappa=-1$). For this case we conclude that an horizon at the value of the radial coordinate for which the numerator of $\dot{r}_{\kappa=-1}$ is null. Furthermore we note that, due to the denominator of $\dot{r}_{\kappa=-1}$ being positive for $r>r_{\mathrm{div}}$ and the term $\kappa C_f^2/E^2 r^{2p-1}=-C_f^2\,r^{2p-1}/m^2<0$ being negative for all values of $r$, the value of the radial coordinate corresponding to the horizon $r=r_{\mathrm{H}}$ is greater than $r_{\mathrm{div}}$~(\ref{r_div})
\be
\left\{\ba{rcl}p&<&1\\1&=&\displaystyle\left(r_{\mathrm{H}}^{p-1}\,(\tilde{C}_A\,r^{p-1}_{\mathrm{H}}+\tilde{\theta})\right)^2+\frac{C_f^2}{m^2}\, r_{\mathrm{H}}^{2p-1}\ea\right.\ \Leftrightarrow\ r_{\mathrm{H}}>r_{\mathrm{div}}\ .
\lb{r_H}
\ee
Although the author failed to find a analytical solution for this equation the previous discussion is enough to conclude that for all allowed solutions and parameter ranges with $p<1$ there exists an horizon for the value of the radial coordinate $r_{\mathrm{H}}$ given by this equation. Hence both the space-time singularity at $r=0$ and the singularity in the particle velocity at $r=r_{\mathrm{div}}$ are inside the horizon and are not observable by an external observer. This is a valid statement both for photons (which are massive due to the Chern-Simons term) and for any other massive particles.

As for the particular case of solution IV with $p>1$ we note that (further considering the redefinition $r\to 1/r$) the ADM signature of the metric for $r>r_H$~(\ref{r_H}) is $\diag(+,-,-)$, hence with
the opposite sign of the original convention. Recalling that at the horizon the metric changes sign~\cite{gravitation}, this is simply
interpreted as that the interior of the horizon for $p>1$ corresponds to the region with $r>r_H$, hence for an external observer in the region $r\in]0,r_H[$ these solutions are interpreted as a dressed point-like singularity at $r=0$ and an horizon at $r=r_H$ such that $r_{\mathrm{div}}>r_H$ and the singularity at $r=+\infty$ are within the region contained by the horizon ($r>r_H$).

Next we compute the mass, the magnetic flux and the angular momentum for the classical solutions obtained.

\subsection{Mass, Angular Momentum and Magnetic Flux\label{sec.mass}}

In this section we derive and analyze the expressions for the mass, angular momentum and magnetic flux for the solutions
computed~(\ref{3sols}). We postpone a interpretation of these results until the next section~\ref{sec.conc} where all the possible cases
are gathered in table~\ref{table.results} and the results obtained are discussed.

We recall that there are several definitions of mass, namely in~\cite{electric} it was computed the ADM mass~\cite{gravitation,ADM,BT}. Adopting this definition of mass, for the metric parameterization~(\ref{ds}), it is obtained
\be
M_{\mathrm{ADM}}=\left.2h'+4\lambda h\phi\phi'+2 \hat{\epsilon}h e^{-b\phi/2}A_\varphi A'_\varphi\right|^{r\to\infty}_{r\to \delta_M}\ ,
\ee
where $\delta_M$ is a cut-off near the singularity (of order of the Planck Length) introduced
to regularize the singularity at the origin maintaining the mass value finite. However
for the magnetic solutions~(\ref{solsB}) the value of the ADM mass is generally complex. We note that the ADM mass corresponds to the  (classical) eigenvalue of the Hamiltonian constraint, hence, generally, aiming at the quantization of the gravitational sector of the theory. This is not the aiming of the present discussion. Instead of the ADM definition of mass we are taking a classical definition of mass that allows for real values to the solutions~(\ref{solsB}). The standard General Relativity definition of mass is the integral of the gravitational mass-energy density $\rho_g$. For a generic Einstein Tensor $G_{\mu\nu}$ the mass-energy density $\rho_g$ and pressure $p_g$ are~\cite{gravitation}
\be 
\rho_g=G_{00}-p_g(1-g_{00})\ \ \ ,\ \ \ p_g=-\frac{G_{03}}{g_{03}}\ ,
\ee
such that the total mass and angular momentum are obtained by integrating these quantities over a spatial hyper-surface~\cite{gravitation}
\be
M=\int \sqrt{|h_{ij}|}\,\rho_g\,dx^2\ \ \ ,\ \ \ S_z=\int \sqrt{|h_{ij}|}\,r\,p_g\,{g^{03}}\,dx^2\ ,
\ee
where $|h_{ij}|$ stands for the determinant of the induced 2-dimensional spatial metric discussed in the previous section and we
note that in $2+1$-dimensions the only angular momentum component correspond to the $3+1$-dimensional angular momentum along $z$ (from the definition $S_k=\int\epsilon_{kij}x^iT^{0j}$~\cite{gravitation} it is obtained that $S_r=S_\varphi=0$).

For the action~(\ref{S}) there is also a contribution to the classical gravitational mass due to the dilaton-like scalar field $\phi$. This contribution can be read directly from the Einstein Equations~\cite{electric}
\be
G_{\mu\nu}+\lambda\partial_\mu\phi\partial_\nu\phi-\frac{\lambda}{2}g_{\mu\nu}\partial_\alpha\phi\partial^\alpha\phi+\frac{1}{2}e^{b\phi}g_{\mu\nu}\Lambda=2e^{-\frac{b}{2}\phi}T_{\mu\nu}\ ,
\ee
where we have taken in consideration the ansatz $a=0$, $c=-b/2$ and the bare electro-magnetic stress-energy tensor is $T_{\mu\nu}=\hat{\epsilon}\left(F_{\mu\alpha}F_{\mu}^{\ \alpha}-g_{\mu\nu}F^2/4\right)$. Hence we note that, for a classical configuration
obeying these equations, the Einstein tensor contribution plus the scalar field contribution to the gravitational mass-energy density and pressure matches the respective electromagnetic quantities~\cite{gravitation}
\be
\rho_{grav}=\rho_g+\rho_\phi=\rho_{EM}\ \ \ ,\ \ \ p_{grav}=p_g+p_\phi=p_{EM}\ .
\ee
In the following we employ these definitions of gravitational energy-momentum density
and pressure density to compute the respective total quantities.
Noting that the only non-null component of the Maxwell tensor is $F_{r\varphi}=F_{12}=\tilde{B}_*$~(\ref{solsB}) it is straight forward to obtain the expressions
for these quantities
\be
\ba{rcl}
\rho_{grav}=\rho_{EM}&=&\displaystyle-\frac{\hat{\epsilon}}{2}g_{00}g^{11}g^{22}\tilde{B}_*^2\,e^{-\frac{b}{2}\phi}-p_{EM}(1-g_{00})\\[5mm]
&=&\displaystyle-\frac{\hat{\epsilon}\,C_B^2\,C_\phi}{2C_h^2}r^{2p-4}\ ,\\[5mm]
p_{grav}=p_{EM}&=&\displaystyle\frac{\hat{\epsilon}}{2}g^{11}g^{22}\tilde{B}_*^2\,e^{-\frac{b}{2}\phi}\\[5mm]
&=&\displaystyle\frac{\hat{\epsilon}C_B^2\,C_\phi}{2C_h^2}r^{2p-4}\ .\\[5mm]
\ea
\lb{rho_EM}
\ee
These quantities are real valued for all the range of the parameter $p$ and in the limit $p\to 0$ are consistently null, as already discussed the particular solution corresponding to $p=0$ corresponds to Minkowski flat empty space-time. We also note that the equation of state for these solutions is a constant $\omega_{grav}=\rho_{grav}/p_{grav}=-1$. However, depending on the value of the parameter $p$ they may have either a divergence at the origin $r\to 0$ (IR), either a divergence at spatial infinity $r\to +\infty$ (UV) or both. 

To regularize these divergences and allow for a simpler analysis of the total quantities we consider two cut-offs $\delta_{IR}$ (lower cut-off) and
$\delta_{UV}$ (upper cut-off) which can be taken to $0$ and $+\infty$, respectively. Specifically for the mass $M$ we obtain the following integral expression
\be
\ba{rcl}
M&=&\displaystyle\int_{\delta_{IR}}^{\delta_{UV}}dr\int_{0}^{\varphi_{\mathrm{max}}}d\varphi\sqrt{|h_{ij}|}\,\rho_{grav}\\[5mm]
&=&\displaystyle-\frac{\hat{\epsilon}C_B^2C_\phi\pi}{|C_fC_h|}\,\int_{\delta_{IR}}^{\delta_{UV}}dr\,r^{p-3}\left(1-r^{2p-2}(\tilde{C}_A\,r^{p-1}+\tilde{\theta})^2\right)\ ,
\ea
\lb{M_int}
\ee
and for the angular momentum $S_z$
\be
\ba{rcl}
S_z&=&\displaystyle\int_{\delta_{IR}}^{\delta_{UV}}dr\int_{0}^{\varphi_{\mathrm{max}}}d\varphi\sqrt{|h_{ij}|}\,r\,p_{grav}\,g^{02}\\[5mm]
&=&\displaystyle-\frac{\hat{\epsilon}C_B^2C_\phi\pi}{(C_fC_h)^2}\,\int_{\delta_{IR}}^{\delta_{UV}}dr\,r^{p-3}\left(\tilde{C}_A\,r^{p-1}+\tilde{\theta}\right)\times\\[5mm]
&&\displaystyle\hfill\ \times\left(1-r^{2p-2}(\tilde{C}_A\,r^{p-1}+\tilde{\theta})^2\right)\ .
\ea
\lb{Sz_int}
\ee
We note that these quantities are evaluated in a 2-dimensional spatial hyper-plane, hence $M$ has units of mass over length and $S_z$ of mass
such that when embedded into a 3-dimensional spatial manifold it is further required to integrated over the thickness of the 2-dimensional embedding along the orthogonal direction ($z$) to retrieve the standard 3-dimensional quantities with units of mass and angular momentum, respectively. It is relevant to stress that, as discussed in~\cite{duality}, from a $3+1$-dimensional perspective these computations are valid and consistent only for systems with constant fields along the direction orthogonal to the planar system as it is the case of systems with cylindrical symmetry (for further discussions on embedded $2+1$-dimensional systems see for example~\cite{lemos_01} and~\cite{3D}).

Evaluating the integral expression~(\ref{M_int}) for the Mass $M$ we obtain
\be
\ba{rcl}
p&\neq&\displaystyle-1,\,1,\,\frac{6}{5},\,\frac{4}{3},\,\frac{5}{4},\,2\\[5mm]
M&=&\displaystyle-\frac{\hat{\epsilon}C_B^2C_\phi\pi}{|C_fC_h|}\,\left(\frac{\tilde{C}_A^2}{6-5p}\,r^{5p-6}+\frac{2\tilde{C}_A\tilde{\theta}}{5-4p}\,r^{4p-5}\right.\\[5mm]
&&\displaystyle\left.\hspace{2cm}+\frac{\tilde{\theta}^2}{4-3p}\,r^{3p-4}+\frac{1}{p-2}\,r^{p-2}\right)_{\delta_{IR}}^{\delta_{UV}}\ .
\ea
\lb{M}
\ee
For $p=-1$ there are no allowed solution and the specific expressions for $p=1,\, 6/5,\,4/3,\,5/4,\,2$ are listed in appendix~\ref{sec.B} in equations~(\ref{B.M.0}--\ref{B.M.4}).
By direct inspection of the expressions for the mass it is straight forward to conclude that the divergence at the origin $r\to 0$ is present
for $p\leq 2$ and that the divergence at spatial infinity $r\to+\infty$ is present for $p\geq 6/5$. Hence, depending on the
value of the parameter $p$, finite mass expressions $M$ can be evaluated by considering the following limits on $\delta_{IR}$ and $\delta_{UV}$
\be
\ba{rclcl}
p&\in&\displaystyle\left]-\infty,\frac{6}{5}\right[/\left\{-1,0\right\}&\Rightarrow&\left\{\ba{rcl}\delta_{IR}&\not\to&0\\\delta_{UV}&\to&+\infty\ea\right.\\[5mm]
p&\in&\displaystyle\left[\frac{6}{5},2\right]&\Rightarrow&\left\{\ba{rcl}\delta_{IR}&\not\to&0\\\delta_{UV}&\not\to&+\infty\ea\right.\\[5mm]
p&\in&\displaystyle\left]2,+\infty\right]&\Rightarrow&\left\{\ba{rcl}\delta_{IR}&\to&0\\\delta_{UV}&\not\to&+\infty\ea\right.
\ea
\lb{M_limits}
\ee
The first range for the parameter $p$ corresponds to solutions I, II, III and IV with $p\in]2/3,6/5[$~(\ref{3sols}) while the second and third
ranges correspond to solution IV.

As for the sign of the mass, for the range $p\in]-\infty,1]/\{-1,0\}$ it has the opposite sign of $\hat{\epsilon}$, $M\sim-\hat{\epsilon}$ and for the range $p\in]1,+\infty[$ it has the same sign of $\hat{\epsilon}$, $M\sim\hat{\epsilon}$.
We note that a negative mass is not unexpected since we are allowing for a gauge ghost sector, we recall that $\hat{\epsilon}=+1$ corresponds
to a ghost gauge sector and that $\hat{\epsilon}=-1$ corresponds to a standard gauge sector. For the range $p\in]-\infty,1[$, outside the horizon $\rho_{grav}$ has the opposite sign of $\hat{\epsilon}$ in accordance to whether the gauge sector is a ghost or a standard sector, however the predominant contribution to the value of the mass is within the horizon and the integrand in~(\ref{M_int}) changes sign at the horizon such that
the total mass is actually positive when it is considered a ghost gauge sector and it is negative when a standard ghost gauge sector is considered.
in the range $p\in]1,+\infty[$ the opposite behavior is verified such that the total mass is negative when it is considered a ghost gauge sector and it is positive when a standard ghost gauge sector is considered. This is simply explained as due to the contribution of the scalar field to the total mass, its classical energy opposes the contribution from the standard gravitational sector.

Evaluating the integral expression~(\ref{Sz_int}) for the angular momentum $S_z$ we obtain
\be
\ba{rcl}
p&\neq&\displaystyle -1,\,1,\,\frac{7}{6},\,\frac{6}{5},\,\frac{5}{4},\,\frac{4}{3},\,\frac{3}{2},\,2\\[5mm]
S_z&=&\displaystyle-\frac{\hat{\epsilon}C_B^2C_\phi\pi}{(C_fC_h)^2}\,\left(\frac{\tilde{C}_A^3}{7-6p}\,r^{6p-7}+\frac{3\tilde{C}_A^2\tilde{\theta}}{6-5p}\,r^{5p-6}\right.\\[5mm]
&&\displaystyle\hspace{2cm}+\frac{3\tilde{C}_A\tilde{\theta}^2}{5-4p}\,r^{4p-5}+\frac{\tilde{\theta}^3}{4-3p}\,r^{3p-4}\\[5mm]
&&\displaystyle\hspace{2cm}\left.+\frac{\tilde{C}_A}{3-2p}\,r^{2p-3}+\frac{\tilde{\theta}}{p-2}\,r^{p-2}\right)_{\delta_{IR}}^{\delta_{UV}}\ .
\ea
\lb{Sz}
\ee
The specific expressions for $p=-1,\,7/6,\,6/5,\,5/4,\,4/3,\,3/2,\,2$ are listed in appendix~\ref{sec.B} in equations~(\ref{B.Sz.0}--\ref{B.Sz.6}).
By direct inspection of the expressions for the angular momentum it is straight forward to conclude that the divergence at the origin is present
for $p\leq 2$ and that the divergence at spatial infinity is present for $p\geq 7/6$. Hence, depending on the
value of the parameter $p$, finite angular momentum expressions $S_z$ can be evaluated by considering the following limits on $\delta_{IR}$ and $\delta_{UV}$
\be
\ba{rclcl}
p&\in&\displaystyle\left]-\infty,\frac{7}{6}\right[/\left\{-1,0\right\}&\Rightarrow&\left\{\ba{rcl}\delta_{IR}&\not\to&0\\\delta_{UV}&\to&+\infty\ea\right.\\[5mm]
p&\in&\displaystyle\left[\frac{7}{6},2\right]&\Rightarrow&\left\{\ba{rcl}\delta_{IR}&\not\to&0\\\delta_{UV}&\not\to&+\infty\ea\right.\\[5mm]
p&\in&\displaystyle\left]2,+\infty\right]&\Rightarrow&\left\{\ba{rcl}\delta_{IR}&\to&0\\\delta_{UV}&\not\to&+\infty\ea\right.
\ea
\lb{Sz_limits}
\ee
Similarly to the results obtained for the mass, the first range for the parameter $p$ corresponds to solutions I, II, III and IV with $p\in]2/3,7/6[$~(\ref{3sols}) while the second and third ranges correspond to solution IV.

As for the sign of the angular momentum $S_z$ we obtain that in the range $p\in]-\infty,1[/\{-1,0\}$ it is $S_z\sim+\hat{\epsilon}\sign(\tilde{C}_A)$ which correspond to solution III in the range $p\in]-\infty,0[/\{-1\}$, solution I and II in the range $p\in]0,2/3[$ and solution IV in the range $p\in]2/3,1[$. For all these cases $\tilde{C}_A\sim \sign(m)$ such that the sign of the angular momentum is $S_z\sim+\hat{\epsilon}\sign(m)$. For
$p=1$ we obtain that $S_z\sim -\hat{\epsilon}\sign(m)$. In the range $p\in]1,1.2857[$ with $\tilde{\theta}\neq 0$ it is $S_z\sim-\hat{\epsilon}\sign(\tilde{\theta})$ corresponding to the solution IV. When $\tilde{\theta}=0$, in the
range $p\in]1,5/4[$ it is $S_z\sim-\hat{\epsilon}\sign(\tilde{C}_A)$ for which $\tilde{C}_A\sim-\sign(m)$ such that $S_z\sim+\hat{\epsilon}\sign(m)$, for $p=5/4$ it
is $S_z\sim-\hat{\epsilon}\sign(\tilde{C}_A(1-\tilde{C}_A^2))$ for which $\tilde{C}_A=-\sqrt{31}\sign(m)$ such that $\tilde{C}_A(1-\tilde{C}_A^2)=30\sqrt{31}\sign(m)$, hence $S_z\sim-\hat{\epsilon}\sign(m)$ and in the range $p\in]5/4,1.2857[$ it is $S_z\sim+\hat{\epsilon}\sign(\tilde{C}_A)$ for which $\tilde{C}_A=-\sign(m)$ such that $S_z\sim-\hat{\epsilon}\sign(m)$. In the range $p\in[1.2857,+\infty[$ it is
$S_z\sim+\hat{\epsilon}\sign(\tilde{C}_A)$ corresponding to solution IV with $\tilde{C}_A\sim-\sign(m)$, hence we obtain $S_z\sim-\hat{\epsilon}\sign(m)$.

As for the magnetic flux we note that for action~(\ref{S}) the equations of motion are expressed in terms of the covariant electro-magnetic fields ${\mathcal{B}}=\sqrt{-g}e^{c\phi}\tilde{B}_*$ and ${\mathcal{E}}=\sqrt{-g}e^{c\phi}\tilde{E}_*$ instead of the bare electro-magnetic fields $\tilde{B}_*$ and $\tilde{E}_*$~\cite{gravitation}
and that for stationary solutions (not depending explicitly on the time coordinate) the Bianchi identities for the Maxwell tensor can also be re-expressed with respect to these quantities. Hence the Maxwell equations are defined by the covariant fields $\mathcal{B}$ and $\mathcal{E}$
such that the measurable magnetic field is $\mathcal{B}$ and its integral over the 2-dimensional manifold is
\be
\ba{rcl}
\Phi_{\mathcal{B}}&=&\displaystyle\int_{\delta_{IR}}^{\delta_{UV}}dr\int_0^{\varphi_{max}}d\varphi \sqrt{|h_{ij}|}\sqrt{-g}e^{c\phi}\tilde{B}_*\\[5mm]
&=&\displaystyle 2C_BC_\phi C_h^2\pi\int_{\delta_M}^{+\infty}dr\,r^p\left(1-r^{2p-2}\left(\tilde{C}_A\,r^{p-1}+\tilde{\theta}\right)\right)\ .
\ea
\lb{Phi_B_int}
\ee
Evaluating this integral expression we obtain
\be
\ba{rcl}
p&\neq&\displaystyle-1,\,1,\,\frac{1}{3},\,\frac{3}{5}\ ,\\[5mm]
\Phi_{\mathcal{B}}&=&\displaystyle 2C_BC_\phi C_h^2\pi\,\left(\frac{1}{1+p}\,r^{p+1}+\frac{\tilde{\theta}^2}{1-3p}\,r^{3p-1}\right.\\[5mm]
&&\displaystyle\left.\hspace{2cm}+\frac{\tilde{\theta}\tilde{C}_A}{1-2p}\,r^{4p-2}+\frac{\tilde{C}_A^2}{3-5p}\,r^{5p-3}\right)_{\delta_{IR}}^{\delta_{UV}}\ .
\ea
\lb{Phi_B}
\ee
The expressions for the particular values of $p=1,\,1/3,\,3/5$ are listed in appendix~\ref{sec.B} in equations~(\ref{B.Phi_B.0}--\ref{B.Phi_B.2}).
Again, depending on the value of the parameter $p$, finite magnetic flux expressions $\Phi_{\mathcal{B}}$ can be evaluated by considering the following limits on $\delta_{IR}$ and $\delta_{UV}$
\be
\ba{rclcl}
p&\in&\displaystyle\left]-\infty,-1\right[&\Rightarrow&\left\{\ba{rcl}\delta_{IR}&\not\to&0\\\delta_{UV}&\to&+\infty\ea\right.\\[5mm]
p&\in&\displaystyle\left]-1,\frac{3}{5}\right]/\{0\}&\Rightarrow&\left\{\ba{rcl}\delta_{IR}&\not\to&0\\\delta_{UV}&\not\to&+\infty\ea\right.\\[5mm]
p&\in&\displaystyle\left]\frac{3}{5},+\infty\right[&\Rightarrow&\left\{\ba{rcl}\delta_{IR}&\to&0\\\delta_{UV}&\not\to&+\infty\ea\right.
\ea
\lb{Phi_B_limits}
\ee
The first range for the parameter $p$ corresponds to solution III,
the second range to solution I, solution II with $p\in]0,1/3[\cup [1/2,3/5]$ and solution III with $p\in]-1,0[$ while the third range corresponds
to solution II with $p\in]3/5,2/3]$ and solution IV~(\ref{3sols}).

As for the sign of the magnetic flux $\Phi_{\mathcal{B}}$ let us note that the sign of $C_\phi$ and $C_B$ are independent of the specific value of the
parameter $p$. $C_\phi$ is always positive, however from the classical solutions of the equations of motion the sign of $C_B$ is arbitrary, this is
simply understood by noting that, in the absence of a electric field the Einstein equations~(\ref{B.E1}--\ref{B.E4}) only
depend on the square of the magnetic field and that the Maxwell equations~(\ref{B.M1}) and~(\ref{B.M2}) with null electric field $\tilde{E}=0$ are invariant under a change of sign of the magnetic field $\tilde{B}\to-\tilde{B}$. Hence only solutions with both non-null electric and magnetic fields are actually sensitive to the relative electromagnetic fields direction (hence the polarization of the electromagnetic fields), both through the Maxwell equations and the '02' Einstein equation. For the specific expressions of the constants given in~(\ref{constantsB}) the choice of the magnetic field sign can be selected by choosing the sign of the free constant $C_h$ which has no consequences at classical level, hence we will proceed our analysis leaving the sign of $C_B$ unspecified. In the range $p\in]-\infty,1/3[$ the magnetic flux sign is $\Phi_{\mathcal{B}}\sim-\sign(C_B)$, for $p=1/3$ it is $\Phi_{\mathcal{B}}\sim\sign(C_B(1-\tilde{C}_A))$ corresponding to solution I for which $\tilde{C}_A=\sqrt{3}/2$ such that $\Phi_{\mathcal{B}}\sim+\sign(C_B)$, in the range
 $p\in]1/3,1[$ it is $\Phi_{\mathcal{B}}\sim+\sign(C_B)$ and in the range $p\in[1,+\infty[$ it is $\Phi_{\mathcal{B}}\sim-\sign(C_B)$.

Next we gather all the results obtained for the solutions~(\ref{solsB}) and discuss possible interpretations for these configurations.

\setcounter{equation}{0}
\section{Discussion of results\label{sec.conc}}

\subsection{Summary of results}
In this work, based on the space-time duality~(\ref{duality1}) discussed in a previous publication~\cite{duality} and resumed in the introduction we have computed the classical solutions listed in equations~(\ref{solsB}-\ref{3sols}) for the gravitational fields,
a scalar field and the gauge fields of Einstein Maxwell Chern-Simons theory described by action~(\ref{S}) with a non-trivial magnetic field and null electric field.
We have analyzed the space-time singularities of such classical configurations and the curvature values in section~\ref{sec.solutions};
the existence of horizons taking in consideration that no massless photons exist in this theory due to the topological mass for the
photon in section~\ref{sec.horizons}, concluding that a geodesic divergence is present in the interior of the horizon, hence not observable by
an external observer; and in section~\ref{sec.mass} were derived the mass, angular momentum and magnetic flux for such configurations. We summarize all these results in table~\ref{table.results} as
a function of the parameter $p\in]-\infty,+\infty[/\{-1\}$.

In the first column of table~\ref{table.results} are listed the several ranges for the value of the parameter $p$, in the column
labeled $\lim_ {r\to+\infty}\varphi_{\mathrm{max}}$ are listed the asymptotic finite values at spatial infinity of the maximum value for the
coordinate $\varphi$ which simultaneously allow the space-time measure and space measure to have as the asymptotic leading
term (also at spatial infinity) a positive exponent of the radial coordinate, in the columns labeled $M_{\mathrm{div}}$, $S_{z,\mathrm{div}}$
and $\Phi_{B,\mathrm{div}}$ is listed whether the mass is divergent near the origin (IR divergence) or the mass is divergent at spatial
infinity (UV divergence) in accordance to the results obtained in equations~(\ref{M_limits}),~(\ref{Sz_limits}) and~(\ref{Phi_B_limits}),
respectively, in the columns labeled $\sign(M)$, $\sign(S_z)$ and $\sign(\Phi_{\mathcal{B}})$ are listed the sign for these quantities evaluated
from the respective expressions~(\ref{M}),~(\ref{Sz}) and~(\ref{Phi_B}) as well as the particular cases listed in appendix~\ref{sec.B},
in the column labeled $\lim_{r\to+\infty}R$ are listed the asymptotic values of the curvature at spatial infinity obtained by inspection
of the curvature~(\ref{R}) and summarized in~(\ref{Rsols}), in the column labeled "Singularities" are listed the location of the space-time singularities obtained by inspection of the scalar invariant $R_{\mu\nu}R^{\mu\nu}$~(\ref{RR}) and summarized in~(\ref{singularities}), in the column labeled "Horizons" it is listed whether the horizon at $r=0$ and $r=r_H$~(\ref{r_H}) exists according to the discussion in section~\ref{sec.horizons},
in the column labeled "Signature" are listed the ADM signatures for the metric for values of the radial coordinate above the horizon
$r>r_H$~(\ref{r_H}) obtained from inspection of the mapped gravitational fields
$f$, $h$ and $A$ given in~(\ref{fhAtilde}) corresponding to the standard ADM metric parameterization~(\ref{ds}) and finally in the
last column labeled "Solution" are listed the correspondence to the solutions of type I, II, III and IV summarized in equation~(\ref{3sols}) for each of the ranges for the values of the parameter $p$.

\begin{landscape}
{
\begin{table}[ht]
\centering
\begin{tabular}{cccccccccccccc}
\tiny$p$&\tiny$\displaystyle\lim_{\tilde{r}\to\infty}\varphi_{\mathrm{max}}$&\tiny$M_{\mathrm{div}}$&\tiny$S_{z,\mathrm{div}}$&\tiny$\Phi_{{\mathcal{B}},\mathrm{div}}$&\tiny$\sign(M)$&\tiny$\sign(S_z)$&\tiny$\sign(\Phi_{\mathcal{B}})$&\tiny$\displaystyle\lim_{\tilde{r}\to\infty}R$&\tiny Singularities&\tiny Horizon&\tiny Signature, $r>r_H$&\tiny Solution\\
   &\tiny  ~(\ref{phimax_limits})&\tiny  ~(\ref{M_limits})&\tiny  ~(\ref{Sz_limits})&\tiny ~(\ref{Phi_B_limits}) &\tiny     (~\ref{M})    &\tiny ~(\ref{Sz})&\tiny ~(\ref{Phi_B})&\tiny ~(\ref{R})&\tiny ~(\ref{RR})&\tiny ~(\ref{r_H})&\tiny ~(\ref{fhAtilde})&\tiny ~(\ref{3sols})\\[2mm]\hline\\
\tiny$\in\left]-\infty,-1\right[$&\tiny --&\tiny IR&\tiny IR&\tiny IR&\tiny $+\hat{\epsilon}$&\tiny $+\hat{\epsilon}\sign(m)$&\tiny \tiny$-\sign(C_B)$&\tiny $0$&\tiny $r=0$&\tiny $\exists_{r=0},\,\exists_{r>0}$&\tiny $(-,+,+)$&\tiny III(Ghost)\\[2mm]
\tiny$\in\left]-1,-\frac{1}{4}\right]$&\tiny --&\tiny IR&\tiny IR&\tiny IR/UV&\tiny $+\hat{\epsilon}$&\tiny $+\hat{\epsilon}\sign(m)$&\tiny \tiny$-\sign(C_B)$&\tiny $0$&\tiny $r=0$&\tiny $\exists_{r=0},\,\exists_{r>0}$&\tiny $(-,+,+)$&\tiny III(Ghost)\\[2mm]
\tiny$\in\left]-\frac{1}{4},0\right[$&\tiny $\frac{1+4p}{C_f}$&\tiny IR&\tiny IR&\tiny IR/UV&\tiny $+\hat{\epsilon}$&\tiny \tiny$+\hat{\epsilon}\sign(m)$&\tiny $-\sign(C_B)$&\tiny $0$&\tiny $r=0$&\tiny $\exists_{r=0},\,\exists_{r>0}$&\tiny $(-,+,+)$&\tiny III(Ghost)\\[2mm]
\tiny$=0$&\tiny --&\tiny --&\tiny --&\tiny --&\tiny --&\tiny --&\tiny --&\tiny --&\tiny --&\tiny --&\tiny empty flat Minkowski\\[2mm]
\tiny$\in\left]0,\frac{1}{3}\right[$&\tiny \tiny $\frac{1+4p}{C_f}$&\tiny IR&\tiny IR&\tiny IR/UV&\tiny $+\hat{\epsilon}$&\tiny $+\hat{\epsilon}\sign(m)$&\tiny $-\sign(C_B)$&\tiny $0$&\tiny $r=0$&\tiny $\exists_{r=0},\,\exists_{r>0}$&\tiny $(-,+,+)$&\tiny I(Ghost) and II\\[2mm]
\tiny$\in\left[\frac{1}{3},\frac{1}{2}\right[$&\tiny $\frac{1+4p}{C_f}$&\tiny IR&\tiny IR&\tiny IR/UV&\tiny $+\hat{\epsilon}$&\tiny $+\hat{\epsilon}\sign(m)$&\tiny $+\sign(C_B)$&\tiny $0$&\tiny $r=0$&\tiny $\exists_{r=0},\,\exists_{r>0}$&\tiny $(-,+,+)$&\tiny I(Ghost)\\[2mm]
\tiny$=\frac{1}{2}$&\tiny --&\tiny =0&\tiny =0&\tiny =0&\tiny $=0$&\tiny $=0$&\tiny $=0$&\tiny $0$&\tiny $r=0$&\tiny $\exists_{r=0},\,\exists_{r>0}$&\tiny $(-,+,+)$&\tiny I(ghost) and II\\[2mm]
\tiny$\in\left]\frac{1}{2},\frac{3}{5}\right]$&\tiny --&\tiny IR&\tiny IR&\tiny IR/UV&\tiny $+\hat{\epsilon}$&\tiny $+\hat{\epsilon}\sign(m)$&\tiny $+\sign(C_B)$&\tiny $0$&\tiny $r=0$&\tiny $\exists_{r=0},\,\exists_{r>0}$&\tiny $(-,+,+)$&\tiny II\\[2mm]
\tiny$\in\left]\frac{3}{5},\frac{2}{3}\right[$&\tiny --&\tiny IR&\tiny IR&\tiny UV&\tiny $+\hat{\epsilon}$&\tiny $+\hat{\epsilon}\sign(m)$&\tiny $+\sign(C_B)$&\tiny $0$&\tiny $r=0$&\tiny $\exists_{r=0},\,\exists_{r>0}$&\tiny $(-,+,+)$&\tiny II\\[2mm]
\tiny$=\frac{2}{3}$&\tiny --&\tiny IR&\tiny IR&\tiny UV&\tiny $+\hat{\epsilon}$&\tiny $+\hat{\epsilon}\sign(m)$&\tiny $+\sign(C_B)$&\tiny $0$&\tiny $r=0$&\tiny $\exists_{r=0},\,\exists_{r>0}$&\tiny $(-,+,+)$&\tiny II and IV\\[2mm]
\tiny$\in\left]\frac{2}{3},1\right[$&\tiny --&\tiny IR&\tiny IR&\tiny UV&\tiny $+\hat{\epsilon}$&\tiny $+\hat{\epsilon}\sign(m)$&\tiny $+\sign(C_B)$&\tiny $0$&\tiny $r=0$&\tiny $\exists_{r=0},\,\exists_{r>0}$&\tiny $(-,+,+)$&\tiny IV\\[2mm]
\tiny$=1$&\tiny --&\tiny IR&\tiny IR&\tiny UV&\tiny $+\hat{\epsilon}$&\tiny $-\hat{\epsilon}\sign(m)$&\tiny $-\sign(C_B)$&\tiny $0$&\tiny $r=0$&\tiny $\exists_{r=0},\,\nexists_{r>0}$&\tiny $(+,-,-)$&\tiny IV\\[2mm]
\tiny$\in\left]1,\frac{7}{6}\right[$&\tiny --&\tiny IR&\tiny IR&\tiny UV&\tiny $-\hat{\epsilon}$&\tiny $-\hat{\epsilon}\sign(\tilde{\theta})$&\tiny $-\sign(C_B)$&\tiny $0$&\tiny $r=0$&\tiny $\exists_{r=0},\,\nexists_{r>0}$&\tiny $(+,-,-)$&\tiny IV\\[2mm]
\tiny$\in\left[\frac{7}{6},\frac{6}{5}\right[$&\tiny --&\tiny IR&\tiny IR/UV&\tiny UV&\tiny $-\hat{\epsilon}$&\tiny $-\hat{\epsilon}\sign(\tilde{\theta})$&\tiny $-\sign(C_B)$&\tiny $0$&\tiny $r=0$&\tiny $\exists_{r=0},\,\nexists_{r>0}$&\tiny $(+,-,-)$&\tiny IV\\[2mm]
\tiny$\in\left[\frac{6}{5},1.2857\right[$&\tiny --&\tiny IR/UV&\tiny IR/UV&\tiny UV&\tiny $-\hat{\epsilon}$&\tiny $-\hat{\epsilon}\sign(\tilde{\theta})$&\tiny $-\sign(C_B)$&\tiny $0$&\tiny $r=0$&\tiny $\exists_{r=0},\,\nexists_{r>0}$&\tiny $(+,-,-)$&\tiny IV\\[2mm]
\tiny$\in\left[1.2857,\frac{3}{2}\right[$&\tiny --&\tiny IR/UV&\tiny IR/UV&\tiny UV&\tiny $-\hat{\epsilon}$&\tiny $-\hat{\epsilon}\sign(m)$&\tiny $-\sign(C_B)$&\tiny $0$&\tiny $r=0$&\tiny $\exists_{r=0},\,\nexists_{r>0}$&\tiny $(+,-,-)$&\tiny IV\\[2mm]
\tiny$=\frac{3}{2}$&\tiny --&\tiny IR/UV&\tiny IR/UV&\tiny UV&\tiny $-\hat{\epsilon}$&\tiny $-\hat{\epsilon}\sign(m)$&\tiny $-\sign(C_B)$&\tiny $\frac{\tilde{C}_A^2}{8}$&\tiny $r=0$&\tiny $\exists_{r=0},\,\nexists_{r>0}$&\tiny $(+,-,-)$&\tiny IV\\[2mm]
\tiny$\in\left]\frac{3}{2},2\right]$&\tiny --&\tiny IR/UV&\tiny IR/UV&\tiny UV&\tiny $-\hat{\epsilon}$&\tiny $-\hat{\epsilon}\sign(m)$&\tiny $-\sign(C_B)$&\tiny $+\infty$&\tiny $r=0,+\infty$&\tiny $\exists_{r=0},\,\nexists_{r>0}$&\tiny $(+,-,-)$&\tiny IV\\[2mm]
\tiny$\in\left]2,+\infty\right[$&\tiny --&\tiny UV&\tiny UV&\tiny UV&\tiny $-\hat{\epsilon}$&\tiny $-\hat{\epsilon}\sign(m)$&\tiny $-\sign(C_B)$&\tiny $+\infty$&\tiny $r=0,+\infty$&\tiny $\exists_{r=0},\,\nexists_{r>0}$&\tiny $(+,-,-)$&\tiny IV\\[2mm]\hline
\end{tabular}
\caption{Resume of solutions as a function of the parameter $p$.\label{table.results}}
\end{table}
}
\end{landscape}

\subsection{Conclusions}

Given the solutions summarized in table~\ref{table.results} we proceed to
interpret them physically. Of particular relevance are the divergences of the physical properties
of the classical configurations, namely the total mass $M$, the total angular momentum $J_z$
and the total magnetic flux $\Phi_{\mathcal{B}}$. A divergence near the space-time singularity
(or singularities) is non uncommon in $2+1$-dimensional space-times, this is mainly
due to that a gravitational potential proportional to $\sim 1/r$ only in $3+1$-dimensional
space-times corresponds to a finite gravitational mass. Also we
note that such a divergence near the singularity is usually associated with a breakdown
of the theory such that a more complete theory is required. A simple regularization
for the divergent quantities is to consider a lower cut-off $\delta_{IR}$ of the order of the
Planck length near the singularity as was considered in~\cite{electric}.

As for configurations for which the total mass $M$, the total angular momentum $J_z$
and the total magnetic flux $\Phi_{\mathcal{B}}$ are divergence when the integral of the respective
densities is considered up to spatial infinity, let us note that considering
a upper cut-off $\delta_{UV}$ for large values of the radial coordinate $r$ is simply interpreted as a description
of a finite size system such that the cut-off $\delta_{UV}$ is
interpreted as the maximum size of the system. Otherwise, for infinite size systems,
it is not mandatory that these quantities be finite, instead they may be interpreted
as cosmological-like solutions for $2+1$-dimensional space-times as long as the respective
densities are finite away from the singularity at the origin. Let us note that even for
a uniformly distributed (meaning constant) mass-energy density in flat Minkowski space-time
we would obtain a divergent total mass when integrating over all space up to spatial infinity.

Hence for the classical configurations discussed here, to regularize the divergence
at the origin for $M$, $S_z$ or $\Phi_{\mathcal{B}}$ we consider the lower cut-off $\delta_{IR}$ to be of the order
of the Planck length $l_p$. To interpret the divergence and the respective upper cut-off $\delta_{UV}$ for large $r$
let us consider three possible cases:
\begin{itemize}
\item \textit{string-like configurations}: a $2+1$-dimensional point-like effective description of matter centered at the origin generating
a magnetic field of finite flux. When embedded into a $3+1$-dimensional space-time with cylindrical symmetry is interpreted as a magnetic string configuration. These configurations should also have a finite mass and finite angular momentum such that the upper cut-off $\delta_{UV}$ is not required;
\item \textit{configurations driven by an external magnetic field}: the upper cut-off $\delta_{UV}$ is justified by the finite range of the applied external field. Hence, from the point of view of $3+1$-dimensions the magnetic field has cylindrical symmetric and is applied orthogonally
to the planar system in the region $r<\delta_{UV}$;
\item \textit{cosmological-like solutions}: an infinite configuration with background magnetic fields such that are allowed total infinite magnetic flux, infinite mass and angular momentum as long as the respective densities are (locally) finite everywhere except at the space-time singularities.
\end{itemize}

By inspection of the table~\ref{table.results} we conclude that, considering only the cut-off $\delta_{IR}$ the solutions with a magnetic field generating a finite total flux, hence being interpreted as a magnetic string-like configuration in an infinite space-time
are achievable only for the parameter range $p\in]-\infty,-1[$ corresponding to solution~III describing ghost gauge fields. For these configurations
also the total mass and total angular momentum are finite. We remark that due to the particular value of the parameter $p=-1$ not allowing for a solution of the equations of motion, this configurations cannot be obtained from flat Minkowski space-time by continuously changing the parameter $p$.

As for the range $p\in]-1,1[$ (considering the lower cut-off $\delta_{IR}$), $M$ and $J_z$
are finite. However, although the magnetic field $\mathcal{B}$ is finite, the total magnetic flux $\Phi_{\mathcal{B}}$ is divergent when
integrating the magnetic field up to spatial infinity, hence these solutions can be interpreted either
as driven by a cylindrical external magnetic field orthogonal to the planar system ranging from the origin up to the upper
cut-off $r<\delta_{UV}$, either as a cosmological-like solution. In addition we note that, when
considering an external magnetic field, the value of the field $\mathcal{B}$ is null for $p=0$ and $p=1/2$. Hence
the solutions corresponding to these values of the parameters are interpreted as two possible backgrounds
upon which the external magnetic field is applied to. Specifically $p=0$ corresponds to empty flat Minkowski, such that
when the magnetic field is turn on the solutions can be changed smoothly and continuously by varying the parameter $p$ (the variation of
the field solutions with the parameter $p$ are continuous and their derivatives with respect to $p$ are also continuous)
describing the deformation induced by the magnetic field, in the range $p\in]-1/4,0[$ corresponding to
solution~III~(\ref{3sols}) for ghost gauge fields, in the range $p\in]0,1/2[$ also for ghost gauge
fields corresponding to solution~I and in the range $p\in]0,1/3[$ for standard gauge fields corresponding
to solution~II. For $p=1/2$ the background corresponds to a neutral dilatonic-like background and the solutions can be
changed smoothly and continuously by varying the parameter
$p$ in the range $p\in]0,1/2[$ describing ghost gauge fields corresponding to solution~I and
in the range $p\in]1/2,2/3]$ describing standard gauge fields corresponding to solution~II. In the range $p\in[2/3,1[$
corresponding to solution~IV describing standard gauge fields the solutions
can also be changed smoothly and continuously by varying the parameter $p$, however when crossing the value $p=2/3$ the
derivative of the field solutions is not continuous such that this range cannot be obtained smoothly by varying the value of the parameter $p$
starting at any of the neutral backgrounds $p=0$ or $p=1/2$.

For values of the parameter $p\in [1,3/2]$ corresponding to solution~IV describing standard gauge fields the metric ADM signature
for values of the radial coordinate above the value of the radial coordinate of the horizon, $r>r_H$, is the opposite to our
original convention, while for the range $r\in]0,r_H[$ the metric has the ADM signature $\diag(-,+,+)$ corresponding to the original convention.
The interpretation for an external observer is that observable space-time is between $r=0$ and the coordinate horizon $r=r_H$~(\ref{r_H}) such that
$r=0$ is a dressed singularity ($r=0$ is both a singularity and an horizon) and a cosmological horizon exists at $r=r_H$.
In addition we note that the geodesics divergence analyzed in section~\ref{sec.horizons} located at $r=r_{\mathrm{div}}$~(\ref{r_div}), is now beyond the cosmological horizon, specifically for $p>1$ we obtain that $r_H<r_{\mathrm{div}}$. These configurations may be interpreted
as cosmological-like configurations in $2+1$-dimensions as the mass-energy density, the magnetic field and pressure are finite in between horizons.

As for the range $p\in ]3/2,+\infty[$ we obtain an exotic configuration for which space-time has two singularities at $r=0$ and $r=+\infty$.
In particular for the range $p\in ]2,+\infty[$, $M$, $S_z$ and $\Phi_{\mathcal{B}}$ have no divergence at the origin having only a
divergence at spatial infinity. Hence by considering the map $\hat{r}= 1/r$ we obtain, for our metric ADM signature convention,
a magnetic string-like configuration for standard gauge fields with both a singularity at the origin within the horizon at $\hat{r}_H>\hat{r}_{\mathrm{div}}$ and a dressed singularity at spatial infinity (spatial infinity is itself both a singularity and an horizon).

We resume the main configuration types discussed in table~\ref{table.sols}.
{\small
\begin{table}[ht]
\centering
\begin{tabular}{ccc}
configuration type&$p$&solution\\[2mm]\hline\\
string-like&$\displaystyle\in\left]-\infty,-1\right[$&III (ghost)\\[5mm]
driven by $B_*$&$\displaystyle\in\left]-1,\frac{1}{2}\right]$&I(ghost) and III(ghost)\\
& &$p=0\ \Leftrightarrow$ neutral background\\[5mm]
               &$\displaystyle\in\left[0,\frac{1}{2}\right[$&II\\
& &$\displaystyle p=0\ \Leftrightarrow$ neutral background\\[5mm]
               &$\displaystyle\in\left[\frac{1}{2},\frac{2}{3}\right]$&II\\
& &$\displaystyle p=\frac{1}{2}\ \Leftrightarrow$ neutral background\\[5mm]
cosmological-like&$\displaystyle\in\left]\frac{2}{3},\frac{3}{2}\right]$&IV \\[4mm]\hline
\end{tabular}
\caption{Resume of discussed configuration types.\label{table.sols}}
\end{table}
}

As a final remark we note that the magnetic string-like configuration corresponding to solution~III for the range of the parameter $p\in]-\infty,-1[$ describing a ghost gauge sector suggests that, for extended gauge theories containing a ghost gauge sector coupled to
magnetic charge~\cite{U2_1,U2_2}, similar magnetically charged solutions may be computed in $3+1$-dimensions~\cite{4D} and $2+1$-dimensions~\cite{3D}.

\vspace{5mm}\noindent {\large\bf Acknowledgments}\\
This work was supported by Grant SFRH/BPD/34566/2007 up to January 2014 and by project CENTRO-01-0145-FEDER-000014 from August 2017 onwards.

\appendix
\setcounter{equation}{0}
\section{Magnetic Solutions\lb{A.magnetic}}

For completeness, in this appendix we re-derive, directly from the equations of motion for action~(\ref{S}) in the Cartan-frame, the solutions~(\ref{solsB}) obtained in the main text from space-time duality. In form notation the action~(\ref{S}) is  
\bea
S&=&-\int_M\Bigg\{e^{a\phi}\left[\tilde{R}*1+2\lambda\,d\phi\wedge*d\phi\right]-e^{b\phi}\Lambda*1\nonumber\\
&&\hspace{2cm}+\hat{\epsilon}e^{c\phi}\left[\tilde{F}\wedge*\tilde{F}+*J\wedge \tilde{A}\right]+\hat{\epsilon}\frac{m}{2}\,\tilde{A}\wedge \tilde{F}\Bigg\}
\nonumber
\eea
using the metric parameterization~(\ref{gpar})
\bea
d\tilde{s}^2=-\tilde{f}^2(dt+\tilde{A}d\varphi)^2+dr^2+\tilde{h}^2d\varphi^2\ .
\nonumber
\eea
The Cartan triad is then given by
\be
\ba{c}
\ba{rcl}
{\,\rm e}^0=d\theta^0&=&\tilde{f}(dt+\tilde{A}d\varphi)\ ,\\[5mm]
{\,\rm e}^1=d\theta^1&=&dr\ ,\\[5mm]
{\,\rm e}^2=d\theta^2&=&\tilde{h}d\varphi\ ,
\ea\\[15mm]
\ba{lll}
e^0_{\ 0}=\tilde{f}\ ,&e^0_{\ 1}=0\ ,&e^0_{\ 2}=\tilde{f}\tilde{A}\ ,\\[5mm]
e^1_{\ 0}=0\ ,&e^1_{\ 1}=1\ ,&e^1_{\ 2}=0\ ,\\[5mm]
e^2_{\ 0}=0\ ,&e^2_{\ 1}=0\ ,&e^2_{\ 2}=\tilde{h}\ ,
\ea
\ea
\lb{B.e-theta}
\ee
such that the line element in the Cartan-frame is
\be
d\tilde{s}^2=e^ie_i=\eta_{ij}d\theta^id\theta^j=-(d\theta^0)^2+(d\theta^1)^2+(d\theta^2)^2\ ,
\lb{B.Cds}
\ee
The electric field $\tilde{E}_*$ and magnetic field $\tilde{B}_*$ in the coordinate frame are given by
\be
\ba{rcl}
\tilde{E}_*&=&\tilde{E}\,\tilde{f}\ ,\\[5mm]
\tilde{B}_*&=&\tilde{B}\,\tilde{h}-\tilde{E}\,\tilde{f}\,\tilde{A}\ ,
\ea
\lb{B.EB*}
\ee
where $\tilde{E}$ and $\tilde{E}$ are the electromagnetic fields in the Cartan-frame.
We note that the metric parameterization~(\ref{gpar}) allows for the electric
field to be null both in the coordinate frame and in the Cartan-frame, $\tilde{E}=0\Lra \tilde{E}_*=0$. This
parameterization also allows for the Maxwell equations in the Cartan-frame to have purely magnetic
solutions as we will derive next.

Noting that
\be
\ba{rcl}
de^0&=&-\beta e^0\wedge e^1+\gamma e^1\wedge e^2\ ,\\[3mm]
de^2&=&\alpha e^1\wedge e^2\ ,
\ea
\ee
the Equations of motion, connections, curvature and remaining quantities depend only on the combinations
\be
\alpha=\frac{\tilde{h}'}{\tilde{h}}\ \ ,\ \ \beta=\frac{\tilde{f}'}{\tilde{f}}\ \ ,\ \ \ \gamma=\frac{\tilde{f}\,\tilde{A}'}{\tilde{h}}\ .
\ee
The non null connections in the Cartan-frame are
\be
\ba{l}
\omega^0_{\ 10}=\omega^1_{\ 00}=\beta\ ,\\[5mm]
\omega^0_{\ 12}=\omega^1_{\ 02}=\omega^1_{\ 20}=-\omega^0_{\ 21}=-\omega^2_{\ 01}=-\omega^2_{\ 10}=\gamma/2\ ,\\[5mm]
\omega^1_{\ 22}=-\omega^2_{\ 12}=-\alpha\ ,\\[5mm]
\ea
\ee
and the Einstein and the energy-momentum tensor components are
\be
\ba{c}
\ba{rcl}
\tilde{G}_{00}&=&-\alpha^2+3\gamma^2/4-\alpha'\ ,\\[3mm]
\tilde{G}_{11}&=&\alpha\beta+\gamma^2/4\ ,\\[3mm]
\tilde{G}_{22}&=&\beta^2+\gamma^2/4+\beta'\ ,\\[3mm]
\tilde{G}_{02}&=&\beta\gamma+\gamma'/2\ ,
\ea
\ \ \
\ba{rcl}
2\tilde{T}_{00}&=&\hat{\epsilon}\left(\tilde{B}^2+\tilde{E}^2\right)\ ,\\[3mm]
2\tilde{T}_{11}&=&\hat{\epsilon}\left(\tilde{B}^2-\tilde{E}^2\right)\ ,\\[3mm]
2\tilde{T}_{22}&=&\hat{\epsilon}\left(\tilde{B}^2+\tilde{E}^2\right)\ ,\\[3mm]
2\tilde{T}_{02}&=&-2\hat{\epsilon}\tilde{B}\tilde{E}\ ,
\ea
\\[17mm]
\ba{rcl}
\Phi_{00}&=&-a\phi''+(\lambda/2-a^2)(\phi')^2\ ,\\[3mm]
\Phi_{11}&=&\lambda/2(\phi')^2\ ,\\[3mm]
\Phi_{22}&=&a\phi''-(\lambda/2-a^2)(\phi')^2\ .
\ea
\ea
\lb{B.Gij}
\ee
We note that under the duality~(\ref{duality1}) only the dilaton contribution to the energy-momentum tensor is invariant while the Maxwell
energy-momentum tensor acquires a minus sign (this accounts to take $\hat{\epsilon}\to-\hat{\epsilon}$)
and for the Einstein tensor the terms $\gamma^2/4$ and $3\gamma^2/4$ are swapped.
For a direct comparison with the same tensor quantities for the standard metric ADM parameterization~(\ref{ds})
we refer the reader to the appendix of~\cite{electric}). In the following we consider both
cases $\hat{\epsilon}=+1$ and $\hat{\epsilon}=-1$.

The Maxwell Equations are
\bea
\tilde{B}'+\beta \tilde{B}+c\,\tilde{B}\,\phi'&=&\displaystyle m\,\tilde{E}\,e^{-c\phi}\lb{B.M1}\ ,\\[5mm]
\tilde{E}'+\alpha \tilde{E}+c\,\tilde{E}\,\phi'+\gamma \tilde{B}&=&\displaystyle-m\,\tilde{B}\,e^{-c\phi}\ ,\lb{B.M2}
\eea
for purely magnetic solution $\tilde{E}=\tilde{E}*=0$ the Einstein equations are
\bea
\displaystyle\hspace{-2cm} e^{a\phi}\left(\beta\gamma+\frac{\gamma'}{2}\right)&=&0\lb{B.E1}\ ,\\[5mm]
\displaystyle \hspace{-2cm}e^{a\phi}\left[\alpha^2-\frac{3\gamma^2}{4}+\alpha'+a\phi''+\left(a^2-\frac{\lambda}{2}\right)(\phi')^2\right]+\frac{1}{2}e^{b\phi}\Lambda&=&\displaystyle \hat{\epsilon}\tilde{B}^2e^{c\phi}\ ,\lb{B.E2}\\[5mm]
\hspace{-2cm}\displaystyle e^{a\phi}\left[\beta^2+\frac{\gamma^2}{4}+\beta'+a\phi''+\left(a^2-\frac{\lambda}{2}\right)(\phi')^2\right]+\frac{1}{2}e^{b\phi}\Lambda&=&\displaystyle -\hat{\epsilon}\tilde{B}^2e^{c\phi}\ ,\lb{B.E3}\\[5mm]
\hspace{-2cm}\displaystyle e^{a\phi}\left[\alpha\beta+\frac{\gamma^2}{4}+\frac{\lambda}{2}(\phi')^2\right]+\frac{1}{2}e^{b\phi}\Lambda&=&\displaystyle -\hat{\epsilon}\tilde{B}^2e^{c\phi}\ ,\lb{B.E4}
\eea
and the Dilaton equation is
\be
\hspace{-2cm}e^{a\phi}\left[(4a^2-\lambda)\phi''+a\left(4a^2-2\lambda\right)(\phi')^2\right]+(3a-b)e^{b\phi}\Lambda=-\hat{\epsilon}(a+c)\tilde{B}^2e^{c\phi}\ .\lb{B.D}
\ee
From the second Maxwell Equation~(\ref{B.M2}) we obtain
\be
\gamma=-m\,e^{-c\phi}\ .
\lb{B.gam}
\ee
Using~(\ref{B.gam}) in~(\ref{B.E1}) one obtains that $\beta=c \phi'/2$ such that
\be
\tilde{f}=c_f\,e^{\frac{c}{2}\phi}\ ,
\lb{B.fa}
\ee
where $c_f$ is a free integration constant.
From the first Maxwell Equation~(\ref{B.M1}) with $\tilde{E}=0$ we obtain
\be
\tilde{B}=\chi e^{-\frac{3}{2}c\phi}\ ,
\ee
where $\chi$ is an integration constant.
The remain 3 Einstein~(\ref{B.E2}-\ref{B.E4}) are
\bea
\hspace{-2cm}\displaystyle a\phi''+(a^2-\frac{\lambda}{2})(\phi')^2+\alpha^2+\alpha'-\frac{3m^2}{4}e^{-2c\phi}+\frac{1}{2}\Lambda e^{(b-a)\phi}&=&\displaystyle \hat{\epsilon}\chi^2e^{(-a-2c)\phi}\ ,\lb{B.b2}\\[5mm]
\hspace{-2cm}\displaystyle (a+\frac{c}{2})\phi''+(a^2-\frac{\lambda}{2}+\frac{c^2}{4})(\phi')^2+\frac{m^2}{4}e^{-2c\phi}+\frac{1}{2}\Lambda e^{(b-a)\phi}&=&\displaystyle -\hat{\epsilon}\chi^2e^{(-a-2c)\phi}\ ,\lb{B.b3}\\[5mm]
\hspace{-2cm}\displaystyle \frac{\lambda}{2}(\phi')^2+\frac{c}{2}\alpha\phi'+\frac{m^2}{4}e^{-2c\phi}+\frac{1}{2}\Lambda e^{(b-a)\phi}&=&\displaystyle -\hat{\epsilon}\chi^2e^{(-a-2c)\phi}\ ,\lb{B.b4}\\[5mm]
\eea
and Dilaton Equations~(\ref{B.D}) is
\be
\hspace{-2cm}\displaystyle (4a^2-\lambda)\phi''+a(4a^2-2\lambda)(\phi')^2+(3a-b)\Lambda e^{(b-a)\phi}= -\hat{\epsilon}(a+c)\chi^2e^{(-a-2c)\phi}\ .\lb{B.b5}
\ee
Employing the same ansatz of~\cite{electric}
\be
\ba{rcl}
a&=&0\ ,\\[5mm]
c&=&\displaystyle -\frac{b}{2}\ ,\\[5mm]
\lambda&\neq&\displaystyle \frac{b^2}{8}\ ,
\ea
\ee
where the particular case corresponding to  $b^2=8\lambda$ is excluded due to not admitting a solution for the above equations of motion.
Given this ansatz we combine~(\ref{B.b3}) with~(\ref{B.b5}) obtaining
\be
\phi'=\pm\sqrt{c_1}e^{\frac{b}{2}\phi}\ ,
\lb{B.dphi}
\ee
such that the Dilaton is
\be
\phi=-\frac{2}{b}\ln(c_\phi\,r)\ ,
\lb{B.dil}
\ee
where
\be
c_\phi=\frac{|b|}{2}\sqrt{c_1}\ \ ,\ \ c_1=-2\frac{b^2(\hat{\epsilon}\chi^2+2\Lambda)+2\lambda(4\hat{\epsilon}\chi^2+2\Lambda+m^2)}{\lambda(b^2-8\lambda)}\ .
\lb{B.C1a}
\ee
Imposing either of the equations~(\ref{B.b3}) or~(\ref{B.b5}) to be obeyed by this solution we obtain
that
\be
\chi^2=-\hat{\epsilon}\frac{2\Lambda(b^2+12\lambda)+4\lambda m^2}{b^2+24\lambda}\ ,
\lb{B.chi}
\ee
such that $c_1$ is rewritten as
\be
c_1=4\,\frac{m^2-6\Lambda}{b^2+24\lambda}\ ,
\lb{B.C1}
\ee
and from~(\ref{B.b4}) we obtain
\be
\alpha=-\left(16\frac{\lambda}{b^2}+1\right)\frac{1}{2\,r}\ .
\ee
Therefore
\be
\tilde{h}=c_h\,r^{-\frac{8\lambda}{b^2}-\frac{1}{2}}\ ,
\lb{B.h}
\ee
and from~(\ref{B.fa})
\be
\tilde{f}=c_f\,\sqrt{r}\ ,
\lb{B.f}
\ee
where $c_h$ and $c_f$ are free constants.
From~(\ref{B.gam}) we obtain that
\be
\tilde{A}=c_A\,r^{-\frac{8\lambda}{b^2}-1}+c_{A_\infty}\ ,
\ee
where
\be
c_A=\frac{m\,C_h}{C_f\left(\frac{8\lambda}{b^2}+1\right)}\sqrt\frac{1+\frac{24\lambda}{b^2}}{m^2-6\Lambda}\ .
\ee
Replacing these solutions in~(\ref{B.b2}) and demanding this equation to be obeyed we obtain that
\be
\lambda_\pm=\frac{b^2}{8}\,\frac{3\Lambda\mp\sqrt{\Lambda(2m^2-3\Lambda)}}{m^2-6\Lambda}\ .
\lb{B.l}
\ee

It is further required to ensure that all these relations are possible for real valued constants, in particular that $c_1>0$ and $\chi^2>0$.
We note that the condition $c_1>0$ is obeyed in the range $0<\Lambda<m^2/3$ except for the particular case $\Lambda=m^2/6$ for which $c_1=0$. Then,
imposing the condition $\chi^2>0$, we obtain the four possible solutions and respective bounds on the cosmological constant
\be
\hspace{-1cm}\ba{lll}
\left\{\ba{rcl}\hat{\epsilon}&=&+1\\ \lambda&=&\lambda_+\ea\right.\ :&\left\{\ba{rcl}\chi^2&=&\displaystyle\frac{1}{2}\left[-\Lambda+\sqrt{\Lambda(2m^2-3\Lambda)}\right]\\[3mm]                                                c_1&=&\displaystyle\frac{4}{b^2}\left[3\Lambda+m^2+\sqrt{\Lambda(2m^2-3\Lambda)}\right]\\[5mm]&&\displaystyle 0<\Lambda<\frac{m^2}{2}\\[2mm]\ea\right.\\[20mm]

\left\{\ba{rcl}\hat{\epsilon}&=&-1\\ \lambda&=&\lambda_+\ea\right.\ 
:&\left\{\ba{rcl}\chi^2&=&\displaystyle\frac{1}{2}\left[\Lambda-\sqrt{\Lambda(2m^2-3\Lambda)}\right]\\[3mm]
c_1&=&\displaystyle\frac{4}{b^2}\left[3\Lambda+m^2+\sqrt{\Lambda(2m^2-3\Lambda)}\right]\\[5mm]&&\displaystyle 0<\Lambda<\frac{m^2}{6}\,\vee\,\frac{m^2}{2}<\Lambda<\frac{2m^2}{3}\\[2mm]\ea\right.\\[20mm]

\left\{\ba{rcl}\hat{\epsilon}&=&+1\\ \lambda&=&\lambda_-\ea\right.\ 
:&\left\{\ba{rcl}\chi^2&=&\displaystyle\frac{1}{2}\left[\Lambda+\sqrt{\Lambda(2m^2-3\Lambda)}\right]\\[3mm]
c_1&=&\displaystyle\frac{4}{b^2}\left[3\Lambda+m^2-\sqrt{\Lambda(2m^2-3\Lambda)}\right]\\[5mm]&&\displaystyle 0<\Lambda<\frac{m^2}{6}\\[2mm]\ea\right.\\[20mm]

\left\{\ba{rcl}\hat{\epsilon}&=&-1\\ \lambda&=&\lambda_-\ea\right.\ 
:&\left\{\ba{rcl}\chi^2&=&\displaystyle\frac{1}{2}\left[\Lambda+\sqrt{\Lambda(2m^2-3\Lambda)}\right]\\[3mm]
c_1&=&\displaystyle\frac{4}{b^2}\left[3\Lambda+m^2-\sqrt{\Lambda(2m^2-3\Lambda)}\right]\\[5mm]&&\displaystyle \frac{m^2}{6}<\Lambda<\frac{2m^2}{3}\\[2mm]\ea\right.
\ea
\label{A.final_exp}
\ee

\setcounter{equation}{0}
\section{Expressions for $M$, $S_z$ and $\Phi_{\mathcal{B}}$ for particular values of the parameter $p$\label{sec.B}}

In this appendix are listed the explicit expressions for the mass $M$~(\ref{M_int}), angular momentum $S_z$~(\ref{Sz_int})
and magnetic flux $\Phi_{\mathcal{B}}$~(\ref{Phi_B_int}) for the particular values of the parameter $p$ not included in the
expressions~(\ref{M}),~(\ref{Sz}) and~(\ref{Phi_B}).

Evaluating the integral expression for the mass $M$ for $p=1$ with $A$ given in~(\ref{A_p1}) we obtain
\be
\ba{rcl}
p&=&1\ ,\\[5mm]
M&=&\displaystyle-\frac{\hat{\epsilon}C_B^2C_\phi\pi}{|C_fC_h|}\,\left(r^{-1}\,\left(1-\tilde{C}_A^2-(\tilde{C}_A+\tilde{\theta})\times\right.\right.\\[5mm]
&&\displaystyle\hfill\ \left.\left.\times(\tilde{C}_A+\tilde{\theta}+2\tilde{C}_A\,\log(r))-\tilde{C}_A^2\,\log(r)^2\right)\right)_{r=\delta_{IR}}\ ,\\[5mm]
\ea
\lb{B.M.0}
\ee
for $p=6/5$ evaluating~(\ref{M_int}) we obtain
\be
\ba{rcl}
p&=&\displaystyle\frac{6}{5}\ ,\\[5mm]
M&=&\displaystyle\frac{\hat{\epsilon}C_B^2C_\phi\pi}{|C_fC_h|}\,\Bigg(-\frac{5}{4}r^{-\frac{4}{5}}\left(8\tilde{C}_A\tilde{\theta}\,r^\frac{3}{5}+2\tilde{\theta}^2\,r^\frac{2}{5}-1\right)\\[5mm]
&&\displaystyle\hfill +\tilde{C}_A^2\log(r)\Bigg)_{\delta_{IR}}^{\delta_{UV}}\ ,
\ea
\lb{B.M.1}
\ee
for $p=4/3$ we obtain
\be
\ba{rcl}
p&=&\displaystyle\frac{4}{3}\ ,\\[5mm]
M&=&\displaystyle\frac{\hat{\epsilon}C_B^2C_\phi\pi}{2|C_fC_h|}\,\Bigg(3r^{-\frac{2}{3}}\left(\tilde{C}_A^2\,r^\frac{4}{3}+4\tilde{C}_A\tilde{\theta}\,r+1\right)\\[5mm]
& &\displaystyle\hfill\ +2\tilde{\theta}^2\log(r)\Bigg)_{\delta_{IR}}^{\delta_{UV}}\ ,
\ea
\lb{B.M.2}
\ee
for $p=5/4$ we obtain
\be
\ba{rcl}
p&=&\displaystyle\frac{5}{4}\ ,\\[5mm]
M&=&\displaystyle\frac{2\hat{\epsilon}C_B^2C_\phi\pi}{3|C_fC_h|}\,\Bigg(r^{-\frac{3}{4}}\left(6\tilde{C}_A^2\,r-6\tilde{\theta}^2\,r^\frac{1}{2}+2\right)\\[5mm]
&&\displaystyle\hfill+3\tilde{C}_A\tilde{\theta}\log(r)\Bigg)_{\delta_{IR}}^{\delta_{UV}}\ ,
\ea
\lb{B.M.3}
\ee
and for $p=2$ we obtain
\be
\ba{rcl}
p&=&2\\[5mm]
M&=&\displaystyle\frac{\hat{\epsilon}C_B^2C_\phi\pi}{12|C_fC_h|}\,\Bigg(r^2\left(3\tilde{C}_A^2\,r^2+8\tilde{C}_A\tilde{\theta}\,r+6\tilde{\theta}^2\right)\\[5mm]
&&\displaystyle\hfill\ -12\log(r)\Bigg)_{\delta_{IR}}^{\delta_{UV}}\ .
\ea
\lb{B.M.4}
\ee

Evaluating the integral expression for the angular momentum $S_z$ for $p=1$ with $A$ given in~(\ref{A_p1}) we obtain
\be
\ba{rcl}
p&=&1\ ,\\[5mm]
S_z&=&\displaystyle+\frac{\hat{\epsilon}C_B^2C_\phi\pi}{(C_fC_h)^2}\,\Bigg(r^{-1}\,\left((\tilde{C}_A+\tilde{\theta})^2+\tilde{C}_A(5\tilde{C}_A+3\tilde{\theta})-\tilde{\theta}\right.\\[5mm]
&&\displaystyle+\tilde{C}_A\log(r)\left(3(\tilde{C}_A+\tilde{\theta})^2+3\tilde{C}_A-1+\tilde{C}_A\log(r)\times\right.\\[5mm]
&&\displaystyle\hfill\ \left.\left.\times(3(\tilde{C}_A+\tilde{\theta})+\tilde{C}_A\log(r))\right)\right)\Bigg)_{r=\delta_{IR}}\ ,\\[5mm]
\ea
\lb{B.Sz.0}
\ee
for $p=7/6$ evaluating~(\ref{Sz_int}) we obtain
\be
\ba{rcl}
p&=&\displaystyle \frac{7}{6}\ ,\\[5mm]
S_z&=&\displaystyle\frac{\hat{\epsilon}C_B^2C_\phi\pi}{(C_fC_h)^2}\,\Bigg(-18\tilde{C}_A^2\tilde{\theta}r^{-\frac{1}{6}}-9\tilde{C}_A\tilde{\theta}^2\,r^{-\frac{1}{3}}-2\tilde{\theta}^3\,r^{-\frac{1}{2}}\\[5mm]
&&\displaystyle\hfill\ +\frac{3}{2}\tilde{C}_A\,r^{-\frac{2}{3}}+\frac{6}{5}\tilde{\theta}\,r^{-\frac{5}{6}}+\tilde{C}_A^3\log(r)\Bigg)_{\delta_{IR}}^{\delta_{UV}}\ ,
\ea
\lb{B.Sz.1}
\ee
for $p=6/5$ we obtain
\be
\ba{rcl}
p&=&\displaystyle \frac{6}{5}\ ,\\[5mm]
S_z&=&\displaystyle\frac{5\hat{\epsilon}C_B^2C_\phi\pi}{12(C_fC_h)^2}\,\Bigg(12\tilde{C}_A^3\,r^{\frac{1}{5}}-36\tilde{C}_A\tilde{\theta}^2\,r^{-\frac{1}{5}}+4\tilde{C}_A\,r^{-\frac{3}{5}}\\[5mm]
&&\displaystyle\hfill\ +3\tilde{\theta}\,r^{-\frac{4}{5}}-6\tilde{\theta}^3\,r^{-\frac{2}{5}}+\frac{36}{5}\tilde{C}_A^2\tilde{\theta}\log(r)\Bigg)_{\delta_{IR}}^{\delta_{UV}}\ ,
\ea
\lb{B.Sz.2}
\ee
for $p=5/4$ we obtain
\be
\ba{rcl}
p&=&\displaystyle \frac{5}{4}\ ,\\[5mm]
S_z&=&\displaystyle\frac{\hat{\epsilon}C_B^2C_\phi\pi}{3(C_fC_h)^2}\,\Bigg(6\tilde{C}_A^3\,r^{\frac{1}{2}}+6\tilde{C}_A\,r^{-\frac{1}{2}}+4\tilde{\theta}\,r^{-\frac{3}{4}}\\[5mm]
&&\displaystyle\hfill\ -12\tilde{\theta}^3\,r^{-\frac{1}{4}}+36\tilde{C}_A^2\tilde{\theta}\,r^{\frac{1}{4}}+9\tilde{C}_A\tilde{\theta}^2\log(r)\Bigg)_{\delta_{IR}}^{\delta_{UV}}\ ,
\ea
\lb{B.Sz.3}
\ee
and for $p=4/3$ we obtain
\be
\ba{rcl}
p&=&\displaystyle \frac{4}{3}\ ,\\[5mm]
S_z&=&\displaystyle\frac{\hat{\epsilon}C_B^2C_\phi\pi}{2(C_fC_h)^2}\,\Bigg(2\tilde{C}_A^3\,r+18\tilde{C}_A\tilde{\theta}^2\,r^{\frac{1}{3}}+6\tilde{C}_A\,r^{-\frac{1}{3}}\\[5mm]
&&\displaystyle\hfill\ +3\tilde{\theta}\,r^{-\frac{2}{3}}+9\tilde{C}_A^2\tilde{\theta}r^{\frac{2}{3}}+2\tilde{\theta}^3\log(r)\Bigg)_{\delta_{IR}}^{\delta_{UV}}\ ,
\ea
\lb{B.Sz.4}
\ee
for $p=3/2$ we obtain
\be
\ba{rcl}
p&=&\displaystyle \frac{3}{2}\ ,\\[5mm]
S_z&=&\displaystyle\frac{\hat{\epsilon}C_B^2C_\phi\pi}{(C_fC_h)^2}\,\Bigg(\frac{1}{2}\tilde{C}_A^3\,r^{2}+3\tilde{C}_A\tilde{\theta}^2\,r+2\tilde{\theta}\,r^{-\frac{1}{2}}\\
&&\displaystyle\hfill\ +2\tilde{\theta}^3\,r^{\frac{1}{2}}+2\tilde{C}_A^2\tilde{\theta}\,r^{\frac{3}{2}}-\tilde{C}_A\log(r)\Bigg)_{\delta_{IR}}^{\delta_{UV}}\ ,
\ea
\lb{B.Sz.5}
\ee
for $p=2$ we obtain
\be
\ba{rcl}
p&=&\displaystyle 2\ ,\\[5mm]
S_z&=&\displaystyle\frac{\hat{\epsilon}C_B^2C_\phi\pi}{20(C_fC_h)^2}\,\Bigg(4\tilde{C}_A^3\,r^{5}+20\tilde{C}_A\tilde{\theta}^2\,r^{3}-20\tilde{C}_A\,r\\[5mm]
&&\displaystyle\hfill\ +10\tilde{\theta}^3\,r^{2}+15\tilde{C}_A^2\tilde{\theta}\,r^{4}-20\tilde{\theta}\log(r)\Bigg)_{\delta_{IR}}^{\delta_{UV}}\ .
\ea
\lb{B.Sz.6}
\ee

Evaluating the integral expression for the magnetic flux $\Phi_{\mathcal{B}}$ for $p=1$ with $A$ given in~(\ref{A_p1}) we obtain
\be
\ba{rcl}
p&=&1\ ,\\[5mm]
\Phi_{\mathcal{B}}&=&\displaystyle\frac{2C_B^2C_h^2C_\phi\pi}{27}\,\left(r^3\,\left(9-2\tilde{C}_A^2+6\tilde{C}_A\tilde{\theta}-9\tilde{\theta}^2\right.\right.\\[5mm]
&&\displaystyle\hfill\ \left.\left.+3\tilde{C}_A\log(r)\left(2\tilde{C}_A-6\tilde{\theta}-3\tilde{C}_A\log(r)\right)\right)\right)_{r=\delta_{UV}}\ ,\\[5mm]
\ea
\lb{B.Phi_B.0}
\ee
for $p=1/3$ evaluating~(\ref{Phi_B_int}) we obtain
\be
\ba{rcl}
p&=&\displaystyle\frac{1}{3}\ ,\\[5mm]
\Phi_{\mathcal{B}}&=&\displaystyle -C_BC_\phi C_h^2\pi\,\Bigg(-\frac{3}{2}\,r^{\frac{4}{3}}-6\tilde{\theta}\tilde{C}_A\,r^{-\frac{2}{3}}-\frac{3}{2}\tilde{C}_A^2\,r^{\frac{4}{3}}\\[5mm]
&&\displaystyle\hfill\ +2\tilde{\theta}^2\log(r)\Bigg)_{\delta_{IR}}^{\delta_{UV}}\ ,
\ea
\lb{B.Phi_B.1}
\ee
and for $p=3/5$ we obtain
\be
\ba{rcl}
p&=&\displaystyle\frac{3}{5}\ ,\\[5mm]
\Phi_{\mathcal{B}}&=&\displaystyle -C_BC_\phi C_h^2\pi\,\Bigg(-\frac{5}{4}\,r^{\frac{8}{5}}+\frac{5}{2}\tilde{\theta}^2\,r^{\frac{4}{5}}+10\tilde{\theta}\tilde{C}_A\,r^{\frac{2}{5}}\\[5mm]
&&\displaystyle\hfill\ +2\tilde{C}_A^2\log(r)\Bigg)_{\delta_{IR}}^{\delta_{UV}}\ .
\ea
\lb{B.Phi_B.2}
\ee

\end{document}